\documentclass[lettersize,journal]{IEEEtran}
\usepackage{amsmath,amsfonts}
\usepackage{algorithmic}
\usepackage{algorithm}
\usepackage{array}
\usepackage[caption=false,font=normalsize,labelfont=sf,textfont=sf]{subfig}
\usepackage{textcomp}
\usepackage{stfloats}
\usepackage{url}
\usepackage{verbatim}
\usepackage{graphicx}
\usepackage{cite}
\usepackage{color}
\usepackage{kotex}
\usepackage{multirow}
\usepackage{adjustbox}
\usepackage{booktabs}
\newcommand{\etal}{\textit{et al.}}
\usepackage[hidelinks]{hyperref}

\begin{document}

\title{Subjective and Objective Quality Evaluation of Super-Resolution Enhanced Broadcast Images on a Novel SR-IQA Dataset}
\author{
	Yongrok Kim\href{https://orcid.org/0009-0001-1148-4544}{\includegraphics[scale=0.05,trim= -1cm -1cm 0 0]{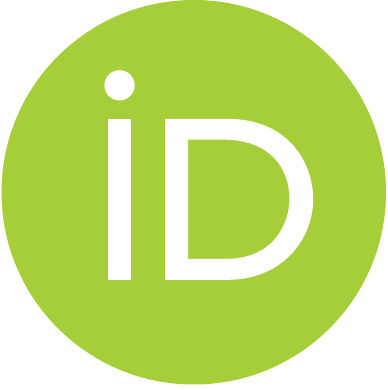}}, 
	Junha Shin\href{https://orcid.org/0009-0009-7556-9983}{\includegraphics[scale=0.05,trim= -1cm -1cm 0 0]{ORCID.pdf}}, 
	Juhyun Lee, 
	Hyunsuk Ko\href{https://orcid.org/0000-0002-7015-8351}{\includegraphics[scale=0.05,trim= -1cm -1cm 0 0]{ORCID.pdf}},~\IEEEmembership{Member,~IEEE}
	\thanks{
		This work was supported by the National Research Foundation of Korea(NRF) grant funded by the Korea government(MSIT) (No. RS-2023-00250751).
		\textit{(Yongrok Kim and Junha Shin contributed equally to this work.)}
		\textit{(Corresponding author: Hyunsuk Ko.)}
	}
	\thanks{This work involved human subjects in its research. The authors confirm that all human subject research procedures and protocols are exempt from Hanyang University Institutional Review Board under Application No. HYUIRB-202102-012}
	\thanks{Yongrok Kim, Junha Shin, and Hyunsuk Ko are with Department of Electrical and Electronic Engineering, Hanyang University ERICA, Ansan 15588, South Korea (e-mail: helloyr12@hanyang.ac.kr; ipip0114@hanyang.ac.kr; hyunsuk@hanyang.ac.kr).}
	\thanks{Juhyun Lee is with Department of Bioengineering, University of Texas at Arlington, Arlington, TX 76019, USA. (e-mail: juhyun.lee@uta.edu).}
}

\maketitle

\begin{abstract}
To display low-quality broadcast content on high-resolution screens in full-screen format, the application of Super-Resolution (SR), a key consumer technology, is essential.
Recently, SR methods have been developed that not only increase resolution while preserving the original image information but also enhance the perceived quality.
However, evaluating the quality of SR images generated from low-quality sources, such as SR-enhanced broadcast content, is challenging due to the need to consider both distortions and improvements.
Additionally, assessing SR image quality without original high-quality sources presents another significant challenge.
Unfortunately, there has been a dearth of research specifically addressing the Image Quality Assessment (IQA) of SR images under these conditions.
In this work, we introduce a new IQA dataset for SR broadcast images in both 2K and 4K resolutions.
We conducted a subjective quality evaluation to obtain the Mean Opinion Score (MOS) for these SR images and performed a comprehensive human study to identify the key factors influencing the perceived quality.
Finally, we evaluated the performance of existing IQA metrics on our dataset.
This study reveals the limitations of current metrics, highlighting the need for a more robust IQA metric that better correlates with the perceived quality of SR images.
\end{abstract}


\section{Introduction}
\label{sec:introduction}

\IEEEPARstart{T}{he} demand for high-resolution content has escalated with advancements in various visual media services, including broadcasting, streaming such as YouTube, and Over The Top (OTT) platforms, paralleled by the increasing resolution of display devices.
Televisions, for instance, have transitioned from 2K to 4K resolution, and currently, there is a rapid proliferation of 8K TVs.
In response to this evolving trend, content providers are not only making efforts to procure new high-resolution contents but also focusing on converting previously acquired low-resolution content to higher resolutions.
In this context, Super-Resolution (SR) technology, which is one of the consumer technologies and reconstructs High-Resolution (HR) images from Low-Resolution (LR) images, has emerged as a key area of research.
With the advancements in deep learning, various deep learning-based SR methods such as SRCNN\cite{SRCNN}, FSRCNN\cite{FSRCNN}, VDSR\cite{VDSR}, SRGAN\cite{SRGAN}, BSRGAN\cite{BSRGAN}, SwinIR\cite{SwinIR} are also being actively proposed.

When upscaling old broadcast content for full-screen display on high-resolution devices, SR technology must address the inherent challenges of enhancing low-quality, low-resolution images.
These images frequently contain significant sensor noise and blur, attributable to the limitations of camera technology during image acquisition.
Applying SR under these conditions often amplifies or extends these distortions, resulting in a SR image of compromised quality.
Recent SR methods in BSRGAN\cite{BSRGAN}, SwinIR\cite{SwinIR} incorporate image enhancement techniques such as deblurring and denoising, aiming to improve perceptual quality in the resulting SR images.
Therefore, in the evaluation of the quality of SR images, it is crucial to evaluate not only the extent of distortion but also the degree of improvement.

\begin{figure}
	\centering
	\subfloat[]{\includegraphics[width=0.95\columnwidth]{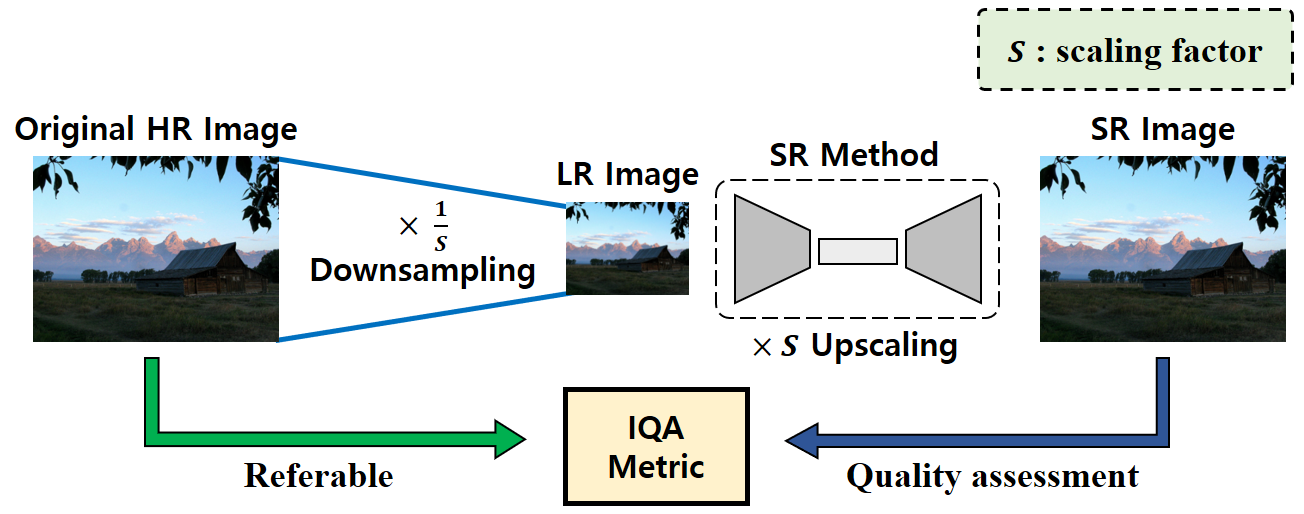}\label{fig:srframework1}} \\ \vspace{-2mm}
	\subfloat[]{\includegraphics[width=0.95\columnwidth]{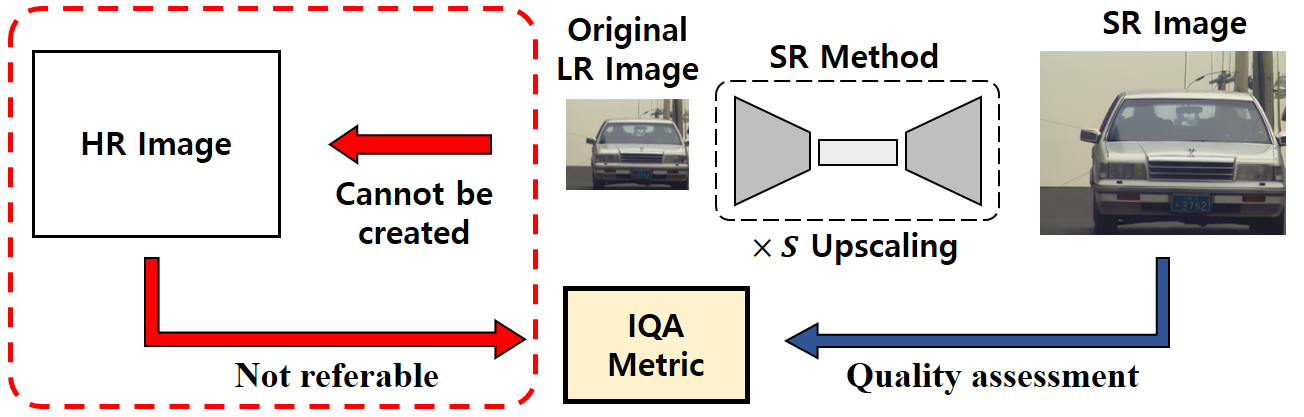}\label{fig:srframework2}}
	\caption{The scenarios of (a) existing SR research and (b) real-world environments on the application of SR and the quality evaluation of SR image.}
	\label{fig:srframework}
	\vspace{-5mm}
\end{figure}

Image Quality Assessment (IQA) is a technology for evaluating images.
It is devided into Full-Reference (FR), Reduced-Reference (RR), and No-Reference (NR) IQA, depending on the availability of reference images.
FR-IQA metrics, such as PSNR, SSIM \cite{ssim}, and LPIPS \cite{lplips}, leverage full reference images to assess image quality.
In contrast, RR-IQA metrics, such as OSVP\cite{osvp}, REDLOG\cite{redlog}, and IGTS\cite{igts}, use reduced reference images like low-resolution images or image patches.
These metrics primarily use high-quality original images as references and assume that these references represent the pinnacle of visual quality.
They evaluate image quality based on the differences between the test image and the reference image, focusing solely on distortions in the test image and not accounting for potential improvements.
On the other hand, NR-IQA metrics such as BRISQUE \cite{brisque}, NIQE \cite{niqe}, and DIIVINE \cite{DIIVINE} evaluate image quality without using reference images.
Instead, they predict a visual quality based on Natural Scene Statistics (NSS) by measuring deviations from the general statistical characteristics observed in natural images.
However, these metrics also face limitations when evaluating SR images that include both distortions and improvements.
To evaluate SR images that encompass both distortions and improvements, new IQA metrics are needed, and an appropriate IQA dataset is essential for developing these metrics.

Most SR-IQA (i.e., IQA for SR images) datasets such as QADS\cite{qads} consist of high-quality original images, their SR counterparts, and corresponding quality scores.
Specifically, LR images are generated by downsampling original images using various scaling factors (e.g., $\times$2, $\times$3, $\times$4).
As shown in Fig.~\ref{fig:srframework}\subref{fig:srframework1}, the SR images are then obtained by upscaling LR images with various SR methods, using the same scaling factor used in the initial downsampling.
This procedure ensures that the SR images match the original images in resolution, thereby enabling the computation of FR IQA metrics.
However, SR images generated with different scaling factors are not derived from the same original image, making fair comparison challenging. 
Additionally, information loss during the downsampling process complicates the analysis of the effects of the SR methods on image quality alone.
Moreover, this experimental setup does not always align with real-world scenarios.
In practical applications, as shown in Fig.~\ref{fig:srframework}\subref{fig:srframework2}, SR methods are applied directly to original LR images without preceding downsampling.
Thus, a reference image of the same resolution does not exist.
To address this issue, Real-SRQ\cite{kltsrq} captured pairs of LR and HR images of the same scene by adjusting the camera's focal length.
However, since Real-SRQ provides high-quality LR and HR images and does not use SR methods that produce SR images with perceptual quality superior to the originals, it is not possible to develop IQA metrics that account for potential improvements in the SR images.

In this paper, we proposed a new SR-IQA dataset specifically for SR images of broadcast content, named Super-Resolution Enhanced Broadcasting contents (SREB) dataset.
Contrary to existing SR-IQA datasets, SREB utilizes the original images in their native resolution, without any prior downsampling.
Moreover, the SR images are designed to consider both distortions and enhancements compared to the original content.
We conducted a subjective quality test using a pairwise comparison method to obtain Mean Opinion Scores (MOSs) for SR images.
The in-depth analysis of the subjective test results was then performed.
Finally, existing IQA metrics were evaluated and compared using our dataset.

The main contributions of this study are outlined as follows: 
\begin{itemize}
	\item \textbf{SREB dataset \& Subjective quality assessment:}
	Our new dataset comprises original images, which are low-quality and low-resolution broadcast content, alongside their SR derivatives, and associated quality scores.
	By applying SR methods that enhance or distort low-quality originals to generate SR images, SREB can be used to develop IQA metrics that account for the distortions and improvements observed in the SR images.
	Additionally, considering real-world scenarios, the original images were used at their native resolution without any prior downsampling.
	For the subjective quality assessment, these images were displayed in their actual size on an Ultra High Definition (UHD) TV.
	A set of 420 subjective quality scores was obtained from 51 participants.
	\item \textbf{Perceptual quality analysis:}
	A human study was conducted to analyze the perceptual quality characteristics of SR images, focusing on the effects of different SR methods and spatial resolutions. 	Additionally, a questionnaire was utilized to investigate the subjective factors perceived by human participants. 
	\item \textbf{Comparative evaluation of existing IQA metrics:}
	We compared the performances of nine NR-IQA metrics and two RR-IQA metrics on our dataset, revealing that the deep learning-based metric, trained on various image degradation expressions, provides superior performance. Furthermore, we analyzed the limitations of applying the existing metrics to SR images of low-quality broadcast content.
\end{itemize}

The rest of the paper is structured as follows. Section~\ref{sec:relatedWork} provides a comprehensive review of the existing literature on SR. In Section~\ref{sec:proposedDataset}, we detail the proposed dataset, while  Section~\ref{sec:subjectiveExperiments} describes the subjective quality assessment conducted. Section~\ref{sec:evaluation} involves a comparative analysis of various image quality models utilizing our dataset.
Finally, concluding remarks are given in Section~\ref{sec:conclusion}.

\section{Related Work}
\label{sec:relatedWork}
The IQA dataset is utilized to evaluate and benchmark the performance of image quality prediction models.
Typically, it comprises a pristine original image along with their distorted counterparts, featuring specific types or degrees of distortion, and subjective image quality scores, MOS, from a subjective quality test.
Notable IQA datasets include LIVE \cite{LIVE}, VDID2014 \cite{VDID2014}, and MARFD \cite{MARFD}.
Some datasets for image quality assessment are tailored to specific scenarios. For example, \cite{camera} uses various camera models and sensor sizes to quantify subjective quality based on objective metrics, \cite{HEVC} adjusts compression parameters to reflect low-bandwidth mobile environments.

\begin{table*}[!t]
	\centering
	\caption{Summary of SR visual quality assessment dataset}
	\label{tab:dataset}
	{\footnotesize
		\begin{adjustbox}{width=1\textwidth}
			\begin{tabular}{ccccccccc}
				\toprule
				Dataset & Year  & Original Scenes & SR Methods & Scaling Factor & \begin{tabular}[c]{@{}c@{}}Original Image\\Resolution\end{tabular} &SR Results & \begin{tabular}[c]{@{}c@{}}Subjective\\Evaluation\end{tabular} & Subjects \\
				\midrule
				Yang \etal \cite{Yang} & 2014  & 10    & 6     & 2, 3, 4 & 481$\times$321 & 540 & ACR & 30 \\
				Yeganeh \etal \cite{Waterloo} & 2015  & 13    & 8     & 2, 4, 8 & 512$\times$512 & 312   & ACR   & 30 \\
				Ma \etal \cite{Ma} & 2016  & 30    & 9     & 2, 3, 4, 5, 6, 8 & 321$\times$481 & 1620  & ACR   & 50 \\
				SISRSet\cite{sisrset} & 2019  & 15    & 8     & 2, 3, 4 & 256$\times$256 - 720$\times$576  & 360   & PC    & 16 \\
				QADS\cite{qads} & 2019  & 20    & 21    & 2, 3, 4 & 500$\times$380 & 980 & PC    & 100 \\
				Real-SRQ\cite{kltsrq} & 2022  & 60    & 10    & 2, 3, 4 & 800$\times$1100 - 2400$\times$2100 & 1,620 & PC    & 60 \\
				SREB  & 2024  & 30    & 7     & 2, 4 & 720$\times$540 & 420   & PC    & 51 \\
				\bottomrule
			\end{tabular}%
		\end{adjustbox}
	}
	\vspace{-5mm}
\end{table*}
Unlike SR datasets, which aim to enhance SR method performance, SR-IQA datasets are designed to evaluate the performance of quality prediction metrics for SR images.
They include various SR images, subjective quality scores, and conditions important for generating SR images, such as scaling factors, SR method types, and image resolutions.
These factors contribute to creating reliable and effective datasets.
Representative SR-IQA datasets include Yang \etal \cite{Yang}, Yeganeh \etal \cite{Waterloo}, Ma \etal \cite{Ma}, SISRSet\cite{sisrset}, QADS\cite{qads}, and Real-SRQ\cite{kltsrq} which are summarized in Table~\ref{tab:dataset}.

\textbf{Yang \etal} used 10 original images from BSD200 to generate 90 LR and 540 SR images, employing 3 scaling factors ($\times$2, $\times$3, $\times$4) and 6 SR methods.

\textbf{Yeganeh \etal} includes 13 original images at resolutions of 512$\times$512, spanning indoor and outdoor scenes, and generated 312 SR images using 3 scaling factors ($\times$2, $\times$4, $\times$8) and 8 interpolation methods.

\textbf{Ma \etal} extended the dataset of Yang \etal~by modifying the LR image generation settings using the BSD200.
There are 180 LR images using 30 original images and 6 scaling factors ($\times$2, $\times$3, $\times$4, $\times$5, $\times$6, $\times$8).
The 1620 SR images were generated using 9 SR methods.

\textbf{SISRSet} comprises 15 original images, sourced from Set5, Set14, and BSD100.
It used bicubic downsampling with 3 scaling factors ($\times$2, $\times$3, $\times$4) to generate LR images, and 8 SR methods were then applied to these LR images, producing a total of 360 SR images.

\textbf{QADS} selected 20 original images from the MDID\cite{mdid} and created 60 LR images from original images, using 3 scaling factors ($\times$2, $\times$3, $\times$4).
It employed 15 traditional and 6 deep learning-based SR methods. Not all methods were applied across all scaling factors, resulting in a total of 980 SR images.

\textbf{Real-SRQ} captured both the original and LR images simultaneously using varying camera focal lengths to produce 60 original images and 180 corresponding LR images across 3 scaling factors.
Applying ten different SR methods, the dataset includes a total of 1620 SR images.

Detailed information on datasets, including SR content, is summarized in Table~\ref{tab:dataset}.

However, existing datasets predominantly generated LR images through downsampling, rather than using raw LR images.
This approach overlooks the practical challenges of dealing with genuinely low-quality, low-resolution images, e.g., found in broadcasting.
Moreover, even for datasets considering LR image generation, such as Real-SRQ\cite{kltsrq}, it is difficult to obtain LR images that accurately fit the original images.
For these reasons, addressing the need for datasets with naturally low-quality LR images is important for developing more robust and practical SR-IQA models.

\section{Proposed Dataset}
\label{sec:proposedDataset}
To reflect real-world scenarios where a full reference image for a SR image does not exist, SR methods should be directly applied to the original LR image without any downsampling.
Furthermore, while conventional SR methods have mainly focused on minimizing distortion that may occur during upsampling process, recent SR studies have made efforts to enhance image quality, where `enhancement' primarily refers to the reduction of distortions present in the original image.
Therefore, SR-IQA dataset needs to be designed to evaluate both distortion and quality enhancements induced by SR process.
To this end, we introduce a new dataset, named SREB, which provides SR images considering the aforementioned aspects and offers subjective quality scores as well obtained through a subjective quality assessment test.

\begin{figure*}[!t]
	\centering
	\includegraphics[width=0.95\linewidth]{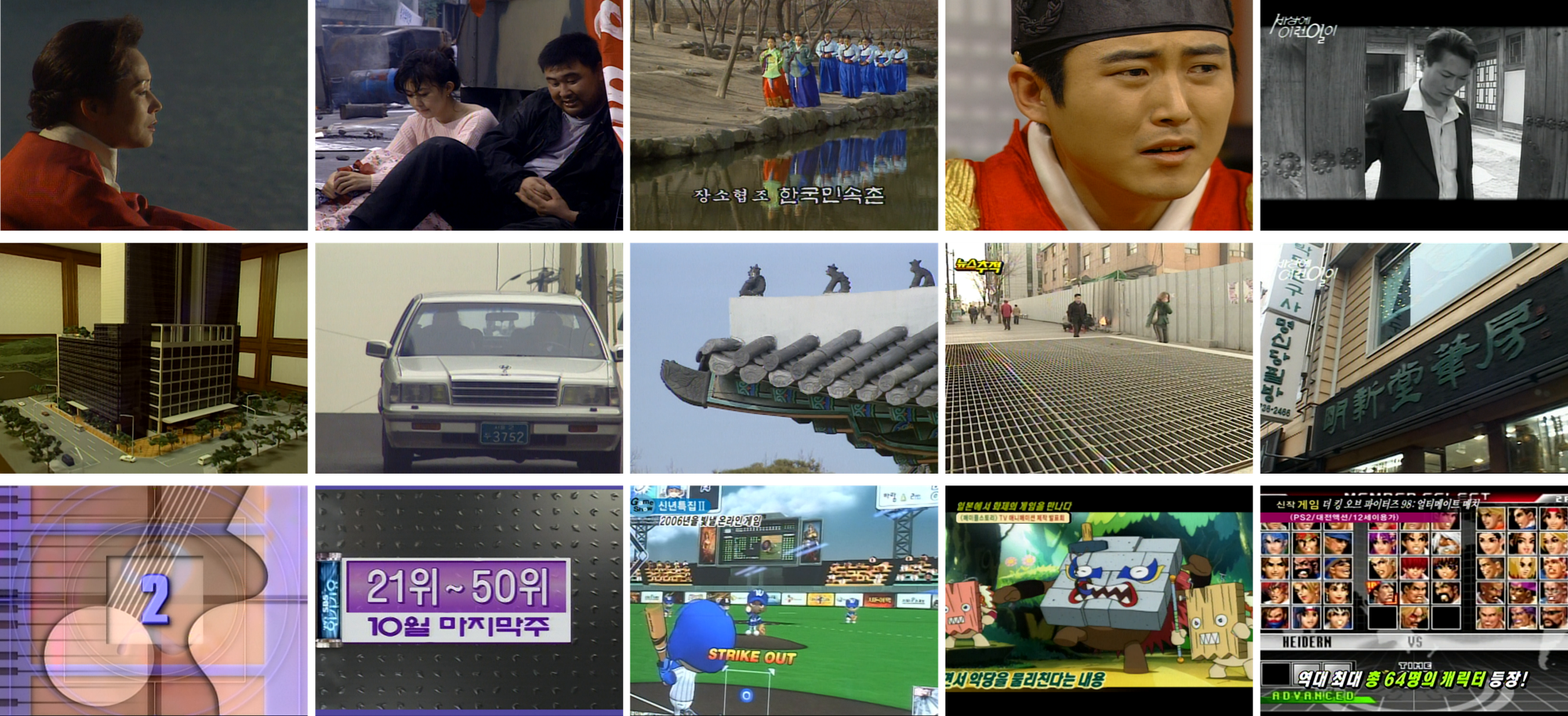}
	\vspace{-2mm}
	\caption{Exemplar original images of our SREB dataset grouped by content characteristics: The images are displayed in a sequence from the first to the third row, each illustrating a distinct content group. Specifically, the image in the first row represents the `Person' group, highlighting human subjects within the dataset. The second row features an image from the `Background' group, showcasing natural or urban environments. Finally, the third row displays an image from the `Screen Content' group, focusing on digital or artificial scenes.}
	\label{fig:our_dataset}
	\vspace{-5mm}
\end{figure*}
\begin{figure}[!t]
	\centering
	\includegraphics[width=0.85\linewidth]{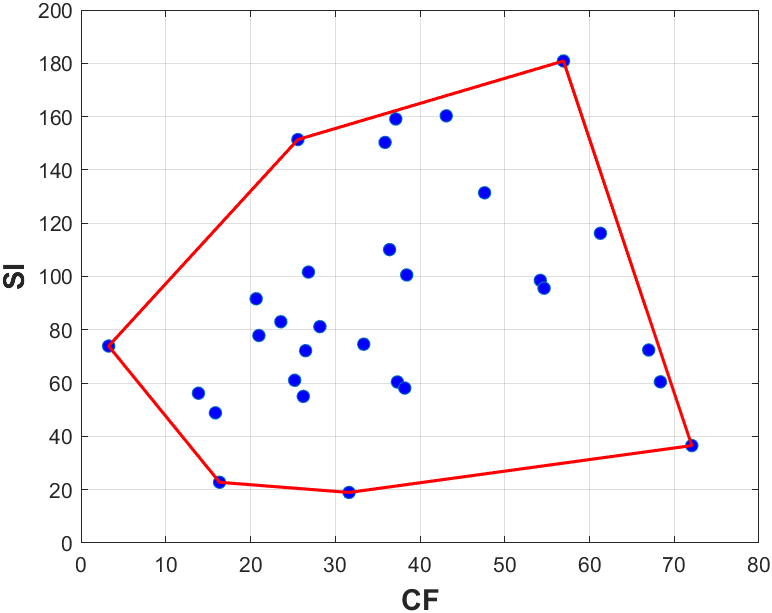}
	\vspace{-2mm}
	\caption{Scatter plot between spatial information (SI) and colorfulness (CF) for the 30 original images in the SREB datase.}
	\label{fig:SICF}
	\vspace{-5mm}
\end{figure}

\textbf{Original images}: 
We attempt a new approach, different from the existing dataset, in constructing an SR-IQA dataset from low-quality original images.
In the real-world scenario, an original image with good cognitive image quality often does not exist due to the limitations of photographing device and applied compression. 
Therefore, it is preferable to use low-quality broadcasting content as the original images in the dataset.
Considering this, the method of obtaining the low-quality original image is as follows.
Original videos were obtained from broadcast  contents of the Korean terrestrial video station SBS between 1995 and 2007, with each video clips being approximately 7-10 seconds long. To take into account a variety of characteristics of broadcast visuals, we divided them into drama, liberal arts, and entertainment categories, selecting a total of 6 program, with two in each category.
The 20 video clips, evenly selected for all content, undergo de-interlacing, cropping, and resizing to convert from an existing 720$\times$512i resolution to a 720$\times$540p resolution. The videos acquired through the process can be analyzed frame by frame. Subsequently, a total of 93 images were selected to cover wide range of the degrees of texture, structure, motion blur and so forth.
To ensure the diversity of scene properties in the SREB dataset, we assessed each image based on spatial information (SI)~\cite{p910} and colorfulness (CF)~\cite{cf}.
These indices were calculated to gauge the spatial complexity and color diversity of image. SI is defined as follows:
\begin{equation}
	SI = \sigma[Sobel(I_F)]
	\label{eq:si}
\end{equation}
where $I_F$ denotes the luminance component of image, $Sobel(\cdot)$ denotes sobel filter, and $\sigma$ represents standard deviation. CF is defined as follows:
\begin{equation}
	rg=I_R-I_G
\end{equation}
\begin{equation}
	yb=\frac{1}{2}(I_R+I_G)-I_B
\end{equation}
\begin{equation}
	\sigma_{rgyb} = \sqrt{\sigma_{rg}^2 + \sigma_{yb}^2}
\end{equation}
\begin{equation}
	\mu_{rgyb} = \sqrt{\mu_{rg}^2 + \mu_{yb}^2}
\end{equation}
\begin{equation}
	CF = \sigma_{rgyb} + 0.3 * \mu_{rgyb}
\end{equation}
where $I_R, I_G, I_B$ denote the red, green, and blue components of the image, respectively, and $\sigma$ and $\mu$ denote the standard deviation and mean values.

According to the calculated SI and CF values, 30 images were selected by excluding those with overlapping characteristics.
For the diversity of content and texture, we categorize the original images into 3 groups: person (row 1), background (row 2), and screen content (row 3), as shown in Fig.~\ref{fig:our_dataset}.
In Fig.~\ref{fig:SICF}, the 30 selected original images encompass a comprehensive range of spatial and color complexity, effectively representing degree of monotony, normality, and complexity.

\begin{figure*}[!t]
	\centering
	\includegraphics[width=1\linewidth]{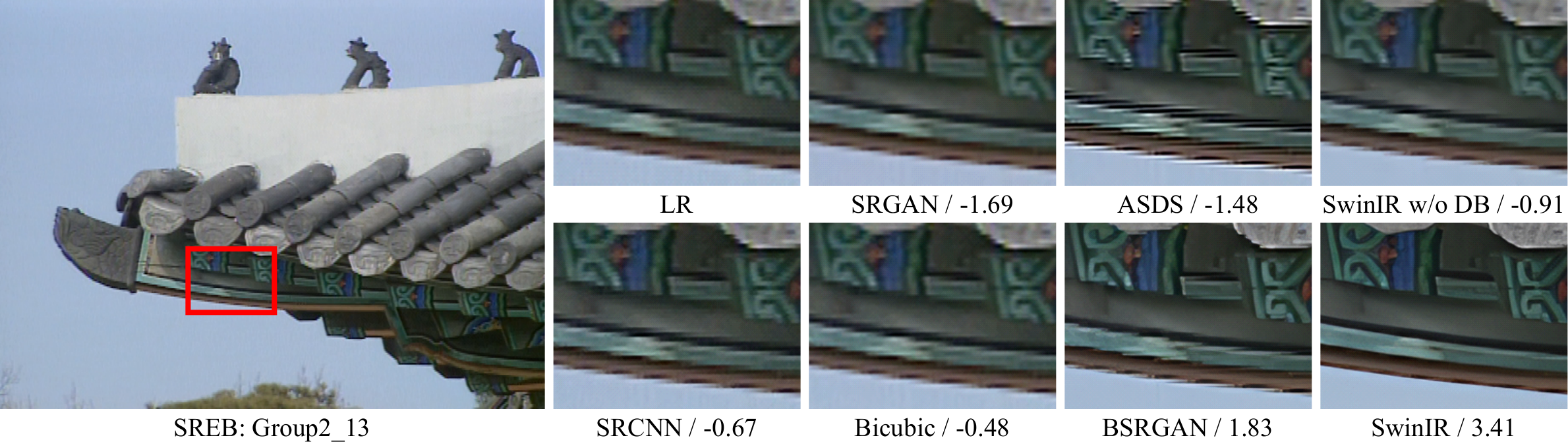}
	\vspace{-6mm}
	\caption{Visual comparison of various SR methods for $\times2$ scaling factor on SREB dataset and MOS corresponding to each SR image.}
	\label{fig:mosimg_1}
	\vspace{-3mm}
\end{figure*}

\begin{figure*}[h]
	\centering
	\includegraphics[width=1\linewidth]{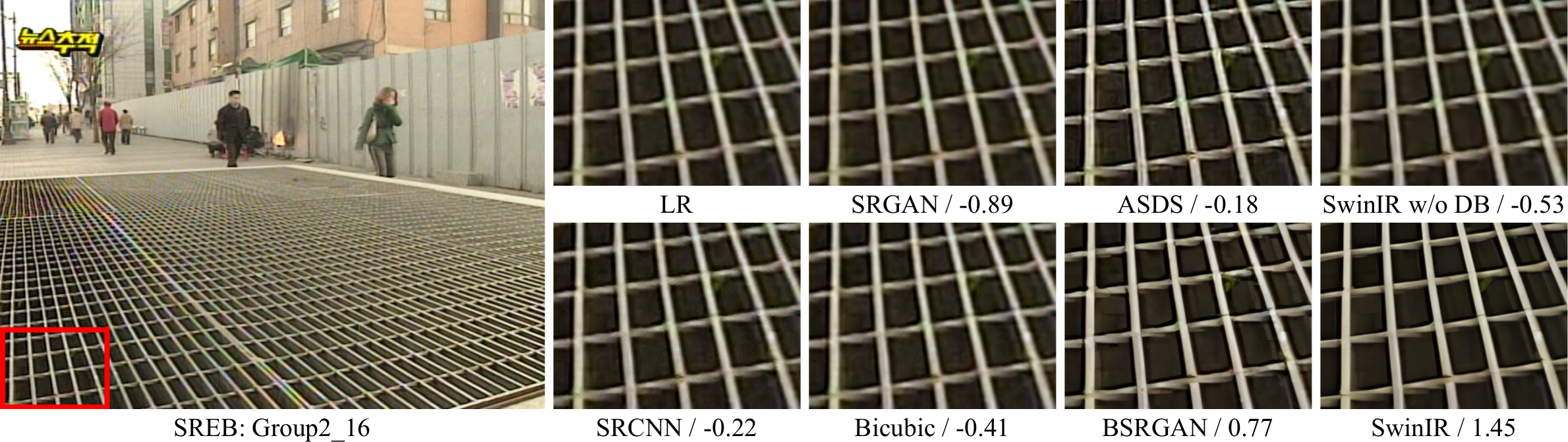}
	\vspace{-6mm}
	\caption{Visual comparison of various SR methods for $\times2$ scaling factor on SREB dataset and MOS corresponding to each SR image.}
	\label{fig:mosimg_2}
	\vspace{-3mm}
\end{figure*}

\textbf{SR images}:
The SR process for the selected 30 original images bypasses downsampling to directly apply SR methods.
This approach ensures that the assessment reflects actual performance without the quality alteration caused by downsampling, simulating real-world application scenarios.
Various types of SR methods, including Bicubic, ASDS\cite{ASDS}, SRCNN\cite{SRCNN}, SRGAN\cite{SRGAN}, BSRGAN\cite{BSRGAN}, SwinIR\cite{SwinIR}, and SwinIR without deblurring (SwinIR w/o DB) were employed.
Here, Bicubic, ASDS, SRCNN, SRGAN, and SwinIR w/o DB exhibit amplified distortion through SR, whereas BSRGAN and SwinIR demonstrate reduced distortion. Existing CNN-based SR methods for SR-IQA datasets have been adopted in various ways, but SR methods by CNN do not show much difference in the results, which may cause confusion in MOS consistency in subjective image quality assessment. Therefore, only one SRCNN representing the CNN-based method is used.
The process of generating the SR image can be defined as follows:
\begin{equation}
	I_{SR}=(I_{LR})\uparrow_{s}
	\label{eq:srimg}
\end{equation}
\begin{equation}
	I_{LR}=I*d_1+d_2+\cdots
	\label{eq:lrimg}
\end{equation}
where $I_{LR}$ is original LR image, $\uparrow_{s}$ denotes SR with a scaling factor of $s$, $I$ is pristine image, and $d_n$ represent different kinds of distortion such as blur and noise that inherently contained in $I_{LR}$. 
In our study, $I_{LR}$ was not synthesized but acquired with distortion included.
The images were processed using scaling factors of $\times$2 and $\times$4, resulting in SR images with resolutions of 1440$\times$1080 and 2880$\times$2160. A total of 420 SR images (30 images$\times$7 methods$\times$2 scaling factors) were generated.
Table~\ref{tab:SRmethod} summarizes the type of SR method, specific method, scaling factor, and the number of images.
Additionally, Fig.~\ref{fig:mosimg_1}, Fig.~\ref{fig:mosimg_2} shows the example of generated SR images along with their respective MOS obtained from the test described in Section~\ref{sec:subjectiveExperiments}.
LR image exhibit blurring and discontinuous edges, which can be either improved or deteriorated by SR.
For instance, SRGAN was evaluated with a low MOS due to the amplification of such distortions, whereas SwinIR received a high MOS as the image quality was enhanced.
\begin{table}[!t]
	\centering
	\caption{Seven SR methods used for generating SR images in the SREB dataset}
	\label{tab:SRmethod}
	{\footnotesize
			\begin{tabular}{ccccc}
				\toprule
				Type								&		Method						&		Year		&		\begin{tabular}[c]{@{}c@{}}Scaling\\ Factor\end{tabular}	&	Images		\\ \midrule
				Interpolation						&		Bicubic						&		- 			&			2, 4			&		60				\\ \hline
				Example based						&		ASDS\cite{ASDS}				& 		2011		&			2, 4			&		60				\\ \hline
				CNN									&		SRCNN\cite{SRCNN}			&		2014		&			2, 4			&		60				\\ \hline
				\multirow{2}{*}{GAN}				&		SRGAN\cite{SRGAN}			&		2016		&			2, 4			&		60				\\ 
				&		BSRGAN\cite{BSRGAN}			&		2021		&			2, 4			&		60				\\ \hline
				\multirow{2}{*}{Transformer}		&		SwinIR\cite{SwinIR}			&		2021		&			2, 4			&		60				\\ 
				&		SwinIR w/o DB	&		2022		&			2, 4			&		60				\\ 
				\bottomrule
			\end{tabular}
		\vspace{-5mm}
	}
\end{table}

\section{Subjective Quality Assessment}
\label{sec:subjectiveExperiments}

\subsection{Subjective Study Design}
\label{sec:experimentsDesign}

In the previous subjective studies on SR images such as \cite{Yang}, \cite{Waterloo}, \cite{Ma}, the Absolute Category Rating (ACR) \cite{p910} was often adopted as the subjective evaluation method. In ACR, the subject views one test image at a time and evaluates its quality using a discrete categorical scale. However, this protocol has limitations in distinguishing subtle quality differences among SR images produced by different SR methods. To address this, recent subjective studies such as \cite{sisrset}, \cite{qads}, \cite{kltsrq} have adopted the Pairwise Comparison (PC) method, as shown in Table \ref{tab:dataset}. This method allows subjects to easily recognize subtle quality difference by selecting the one that they perceive to have superior visual quality from a pair of SR images. Therefore, in this study, we employed the PC method to conduct a subjective quality assessment of SR images. Our evaluation method and environment were established in accordance with ITU-R Rec. BT500-14 \cite{bt500}.

\subsubsection{GUI Configuration}
\label{sec:gui}
\begin{figure*}
	\centering
	\subfloat[]{\includegraphics[height=38.5mm]{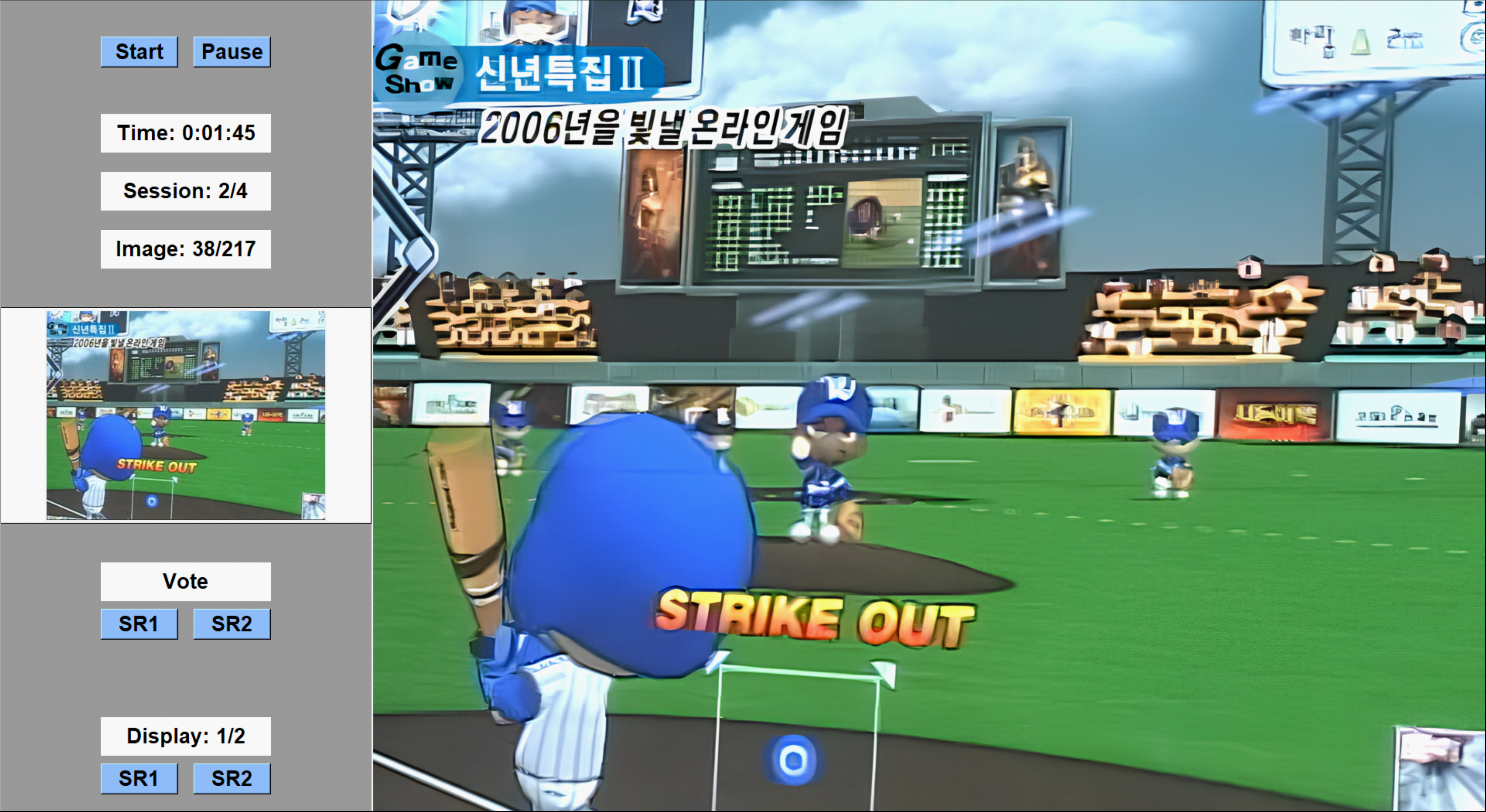}\label{fig:GUI}} \hspace{0.1mm}
	\subfloat[]{\includegraphics[height=38.5mm]{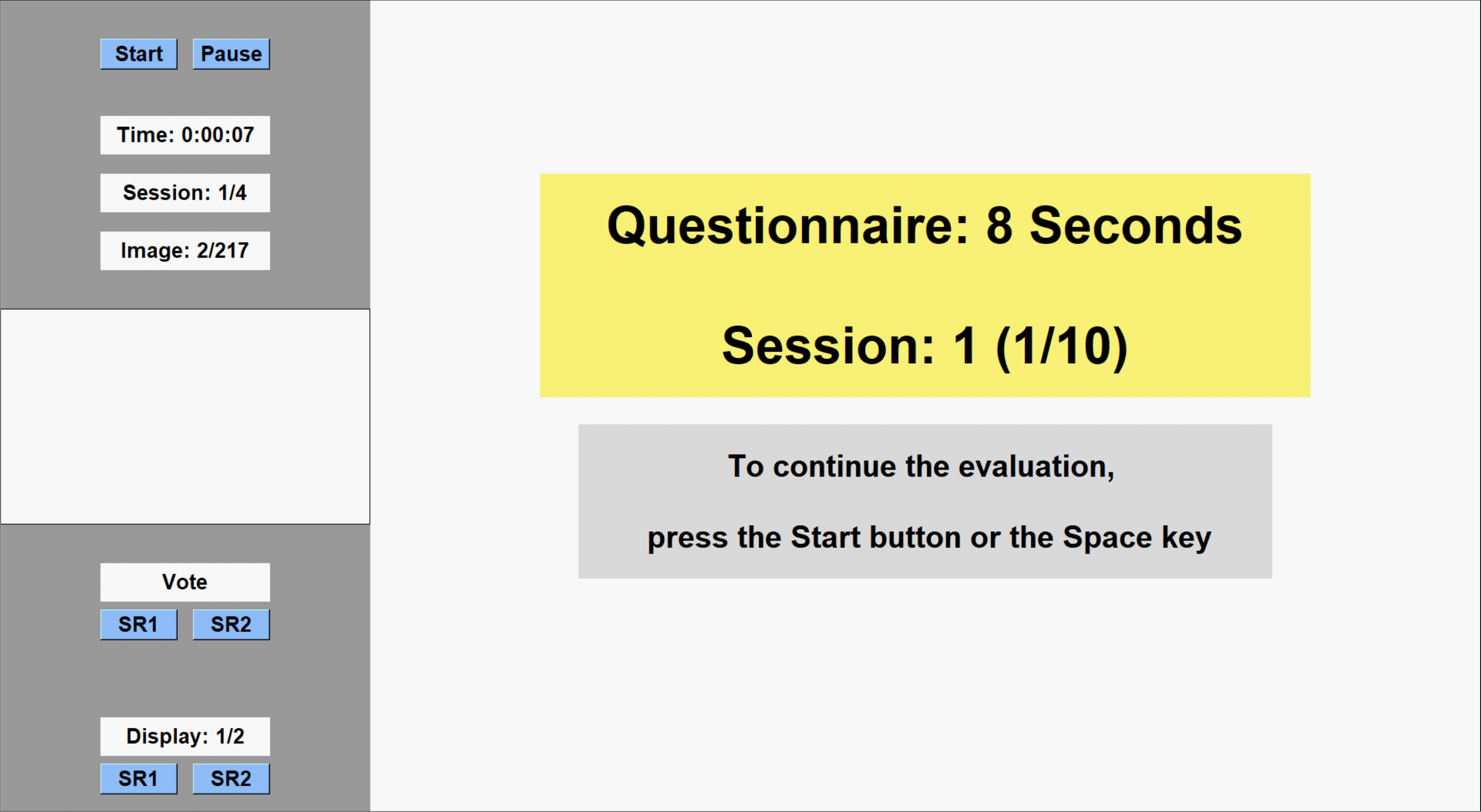}\label{fig:Questionnaire_screen}} \hspace{0.1mm}
	\subfloat[]{\includegraphics[height=38.5mm]{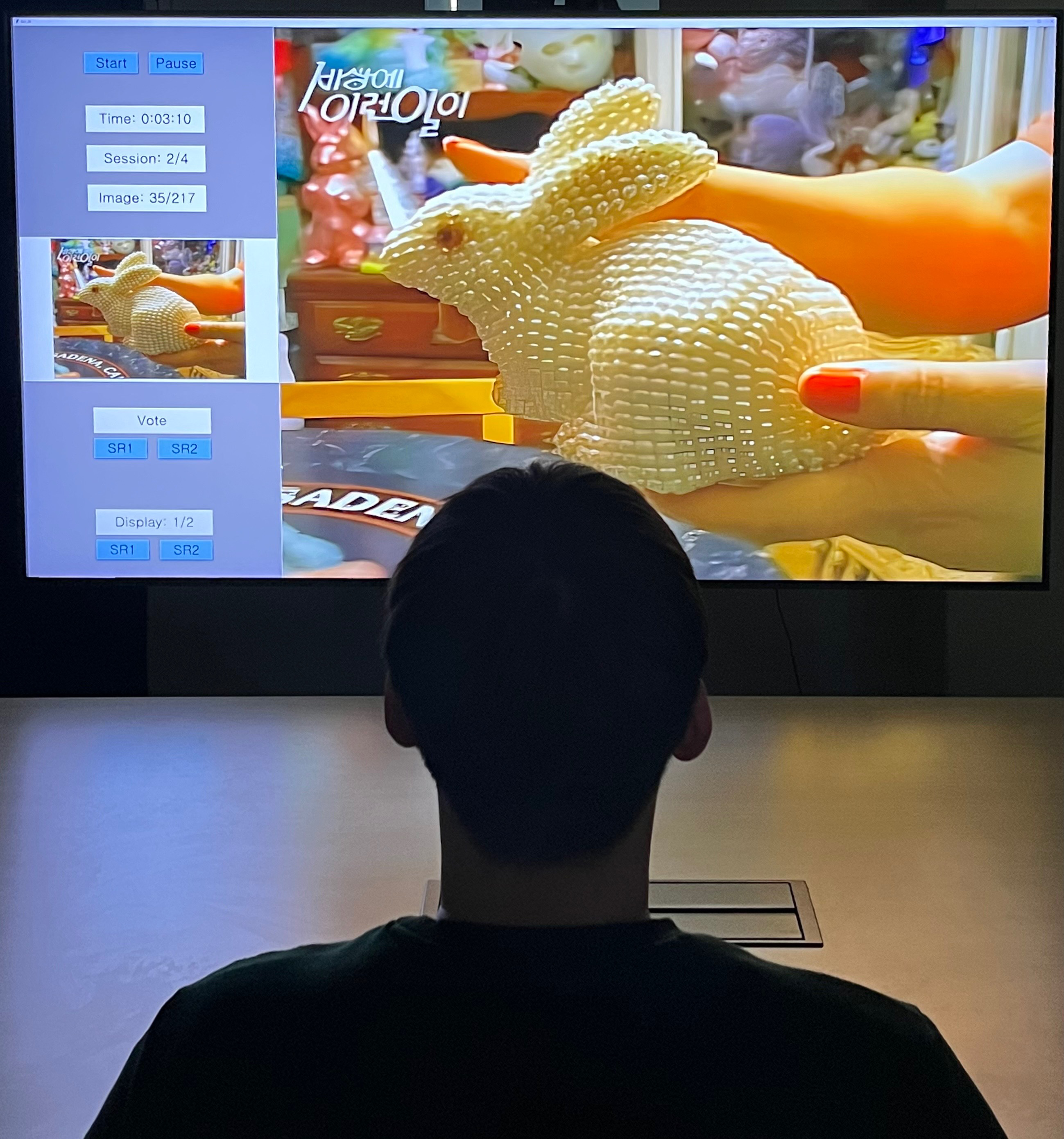}\label{fig:evaluation}}
	\caption{(a) GUI for the evaluation of SR images;
			The top-left section displays time, session, and the number of evaluated images, along with Start/Stop buttons. The middle-left section displays the original LR image corresponding to the SR images. The bottom-left section contains buttons for voting on preferred images and for switching between two SR images for comparison. The right section displays an SR image, which can be switched to another SR image using the display buttons in the bottom-left section.
			(b) Questionnaire screen on the GUI;
			For each survey, a maximum of 8 seconds is given, and the session and survey number are displayed. Subjectors select one primary reason they perceive unselected image to have poor quality from Table \ref{tab:guiquestionnaire} and record it on the provided questionnaire.
			(c) Scene of subjective quality assessment.
	}
	\label{fig:evaluation_view}
	\vspace{-5mm}
\end{figure*}

For our subjective quality test, we implemented a Graphical User Interface (GUI) designed for the PC method, which is shown in Fig.~\ref{fig:evaluation_view}~\subref{fig:GUI}. The GUI comprises four main sections: \\
\textbf{Top-left} displays evaluation progress—test running time, the count of comparison executed, and session information—with start/pause controls.\\
\textbf{Middle-left} shows the original LR image for reference to identify changes in SR images. \\
\textbf{Bottom-left} features voting and display switch buttons for two SR images. \\
\textbf{Right} presents a single SR image, with options to toggle between two SR images for comparison. This facilitates easier comparison for the subject.

Interaction with the GUI is keyboard-friendly, facilitating efficient comparison and selection processes. 
Subjective experiments were conducted on a Samsung 65-inch 4K UHD TV. The viewing distance was set to 1.6 times the display height.
Original LR and corresponding SR images are shown at their actual resolutions, 720$\times$540 for LR, and 1440$\times$1080 ($\times$2) or 2880$\times$2160 ($\times$4) for SR images.
This allowed for considering only the impact of SR methods in subjective image quality assessment with the same scaling factor and enabled the analysis of human perceptual systems based on resolution with different scaling factors.
Fig.~\ref{fig:evaluation_view}~\subref{fig:evaluation} illustrates the evaluation of SR images using our GUI.

\subsubsection{Comparison Pairs and Evaluation Sessions}
\label{sec:Sessions}
To conduct PC-based evaluation, SR images for the same content are paired in sets of two. Each source content was processed using 7 different SR methods with 2 scaling factors, yielding 7 SR images at both $\times$2 and $\times$4 scaling factors. Thus, each source content has 21 pairs (${}_{7}{\rm C}_{2} = 21$) for comparison at each scaling factor. Additionally, 49 pairs ($7 \times 7 = 49$) were generated for cross-scaling factor comparisons ($\times$2 vs. $\times$4). Therefore, for a single source content, there are a total of 91 comparisons pairs ($21 + 21 + 49 = 91$). With 30 source contents in total, the overall number of comparison pairs amounts to 2,730 ($91 \times 30 = 2,730$).

A total of 51 subjects participated in our subjective quality assessment test.
All subjects provided consent for participation in this test.
We divided the 51 participants into three equal groups, with each group assessing SR images for 10 distinct source contents. As a result, each participant conducted a total of 910 votes ($91 \times 10 = 910$). In total, we collected 46,410 votes ($910 \times 51 = 46,410$) from all 51 participants. 

The test was structured into four sessions: the first two sessions focused on comparisons in the same $\times$2 or $\times$4 scaling factor, respectively, while the final two sessions compared images across two scaling factors.
The last two sessions aimed to investigate how scaling factors and resolution differences affect perceived image quality.
The results from the first and second sessions were used to compute MOS for $\times$2 and $\times$4 scaling factors, respectively, while the results from all four sessions were combined to compute MOS for the integrated $\times$2 and $\times$4 scaling factors (i.e., $\times$2$\times$4).
To mitigate visual fatigue, we limited sessions to 30 minutes and incorporated a 5-minute break between each. On average, participants completed the experiment in 98 minutes.

\subsubsection{Survey}
\label{sec:survey}
To further explore the distortion factors in SR images, we conducted a survey during a subjective experiment. Participants were given a questionnaire before they started the evaluation. During the assessment, a questionnaire screen randomly appears on the GUI with a 1/20 chance after each voting session, as illustrated in Fig.~\ref{fig:evaluation_view}~\subref{fig:Questionnaire_screen}. If the questionnaire screen popped up, participants were asked to identify the main reason they found the image quality unsatisfactory for the image they did not choose in the preceding evaluation. This process aimed to pinpoint the primary causes of distortion in the non-selected images. The questionnaire's components are detailed in Table \ref{tab:guiquestionnaire}, with the resolution factors only used in the 3rd and 4th sessions to compare SR images at varying scaling factors.

\begin{table}[!t]
	\centering
	\caption{Questionnaire items for distortion factors in SR images}
	\label{tab:guiquestionnaire}
	{\footnotesize
		\begin{tabular}{ccl}				
			\toprule
			Factors                     & ID & \multicolumn{1}{c}{Reasons}                                        \\ \midrule
			Noise                       & 1     & The image is not clean and contains noise.                          \\ \hline
			Blur                        & 2     & The image is not clear and blurry.        \\ \hline
			Detail                      & 3     & Distorted patterns, shapes, or edges are present.     \\ \hline
			\multirow{2}{*}{\begin{tabular}[c]{@{}c@{}}Color \& \\ Brightness\end{tabular}}      & 4     & There is unnatural color distortion.                                \\ 
			& 5     & Brightness level is either too dark or bright.          \\ \hline
			\multirow{2}{*}{Resolution} & 6     & High resolution causes visual unnaturalness. \\ 
			& 7     & Low resolution causes visual unnaturalness.  \\ 
			\bottomrule
		\end{tabular}
		\vspace{-5mm}
	}
\end{table}

\subsubsection{Checkpoint}
\label{sec:checkpoint}
For the identification and elimination of outliers, we introduced checkpoint comparison pairs in each session. These pairs consist of SR images generated with SwinIR\cite{SwinIR} and their Gaussian-blurred counterparts. This setting ensures clear differences in image quality so as to check whether subjects faithfully participate in subjective experiments and evaluate if the results are reliable. A total of 20 checkpoints have been added, and if an subject provides incorrect responses in two or more checkpoints, they were considered as outliers and excluded from the results. In our test, all 51 participants passed the checkpoints. 

\subsubsection{Training Session}
\label{sec:Trainingsession}
Before the subjective image quality assessment, all participants underwent a training session. This session covered evaluation procedures, GUI navigation, and questionnaire completion. They also  engaged in hands-on practice with the GUI, evaluating dummy SR images produced in the same manner to those in Section~\ref{sec:proposedDataset}, using four source contents not included in the SREB dataset.

\subsection{Post-processing for Subjective Scores}
\label{sec:processing}
The voting results obtained from PC-based subjective test cannot be directly used as subjective quality scores. Thus, MOSs were derived through post-processing methods. We used the Bradley-Terry (B-T) method \cite{btmodel}, a probabilistic model commonly used for PC results. The MOSs were estimated via Maximum Likelihood Estimation (MLE), as described in \cite{kltsrq}.
Specifically, in a pair of $i$ and $j$, $C_{ij}$ represents the number of times $i$ beats $j$, expressed as follows:
\begin{equation}
	C_{ij} = \begin{cases}
		The~number~of~times~i~beats~j, & i \neq j \\
		0, & i = j
	\end{cases}
	\label{eq:prefer}
\end{equation}
In our study, in a pair of two SR images $i$ and $j$, $C_{ij}$ represents the number of times $i$ is selected, derived from the cumulative voting results of all participants. The B-T model is the probability of $i$ beats $j$, expressed as follows:
\begin{equation}
	P_{ij} = \frac{e^{q_{i}}}{e^{q_{i}}+{e^{q_{j}}}}
	\label{eq:bt}
\end{equation}
Here, $Q = \{q_{1}, q_{2}, \cdots, q_{N}\}$ represents estimated subjective quality score, i.e., MOS, of each image, and our goal is to estimate $Q$ using MLE. The likelihood $P(D|Q)$ represents the probability of the voting results $D$ as follows:
\begin{equation}
	P(D|Q)= \overset{N}{\underset{i=1}{\prod}} \overset{N}{\underset{\substack{j=1 \\ j \neq i}}{\prod}} (P_{ij})^{C_{ij}}
	\label{eq:likelihood}
\end{equation}
For the convenience of calculations, negative log likelihood $L(Q)$ can be expressed as Eq.~(\ref{eq:loglikelihood}) and minimized.
\begin{equation}
	L(Q)= -\overset{N}{\underset{i=1}{\sum}} \overset{N}{\underset{\substack{j=1 \\ j \neq i}}{\sum}} C_{ij}\log{(P_{ij})}
	\label{eq:loglikelihood}
\end{equation}
Using condition ${\partial{L(Q)}}/{\partial{q_{k}}}=0$, $Q$ can be derived as in Eq.~(\ref{eq:minimizing}). 
\begin{equation}
	q_{k}^{t+1} = \log{(\frac{\sum_{j=1}^{N}{C_{kj}}}{\sum_{i=1}^{N}{\frac{C_{ki}+C_{ik}}{e^{q_{k}^{t}}+e^{q_{i}^{t}}}}})},~k= 1, 2, \cdots, N
	\label{eq:minimizing}
\end{equation}
where $t$ represents the number of iterations.
By initializing with $Q = \{1, 1, \cdots, 1\}$ and iteratively updating $Q$ based on the voting results until it converges, MOS of each SR image can be obtained.
Subsequently, they were normalized to have a zero mean.

\begin{figure}[!t]
	\centering
	\includegraphics[width=1\linewidth]{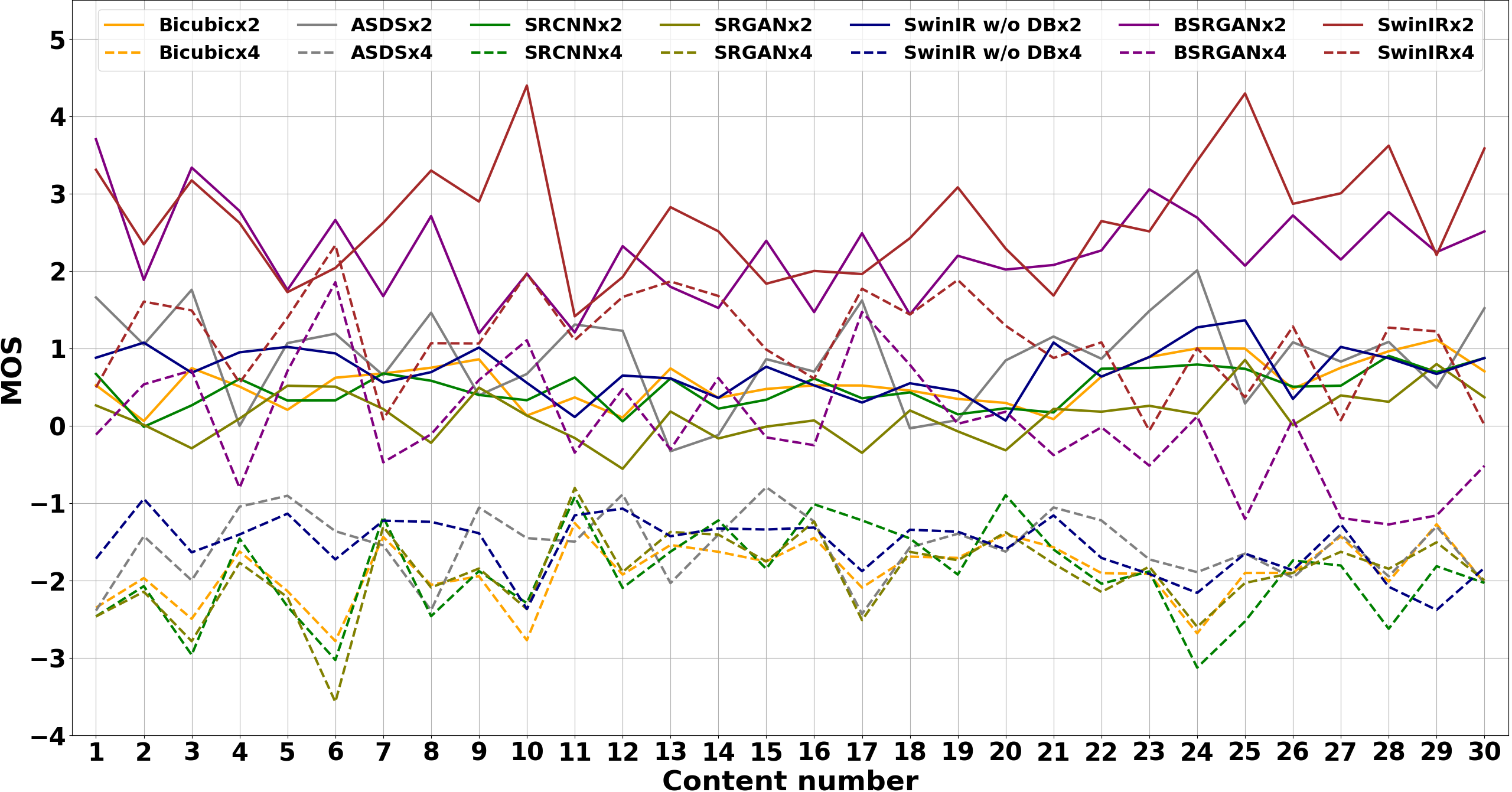}
	\vspace{-6mm}
	\caption{$\times$2$\times$4 MOS score of content by all SR methods. The solid line represents the x2 scale and the dotted line represents the x4 scale.}
	\label{fig:methodscore}
	\vspace{-3mm}
\end{figure}
\begin{table}[!t]
	\centering
	\caption{Summary of the MOS for SR images on SREB dataset}
	\label{tab:mos}
	{\footnotesize
		\begin{tabular}{ccccc}
			\toprule
			Scaling Factor & SR Images & Max  & Min   & Average            \\
			\midrule
			$\times$2                                                       & 210                                                 & 4.61 & -3.27 & \multirow{3}{*}{0} \\ 
			$\times$4                                                       & 210                                                 & 4.44 & -2.35 &                    \\ 
	$\times$2$\times$4                                                      & 420                                                 & 4.39 & -3.57 &                    \\ 
			\bottomrule
		\end{tabular}
	}
	\vspace{-5mm}
\end{table}

\subsection{Study on the Subjective Quality Test Results}
\label{sec:subjectiveStudyResult}
\subsubsection{MOS Analysis}
\label{sec:MOSAnalysis}
The SREB dataset provides MOSs for 420 SR images, with the MOS for each SR image calculated using voting results from the same source content. As a result, 210 MOSs were obtained for of $\times$2 and $\times$4, respectively. In addition, a total of 420 MOSs were further obtained for $\times$2$\times$4. These results are summarized in Table \ref{tab:mos}.

For the comparison of the SR methods and the analysis of the MOS trend by group, the MOS for $\times$2$\times$4 scaling factor of each method for each content is shown in Fig.~\ref{fig:methodscore}.
Each color represents the SR method, with $\times$2 scale represented by a solid line and $\times$4 scale by a dotted line.
The graph shows a consistent trend of better SR quality at lower scaling factors across all methods.
The graph can be divided into four parts:
\begin{itemize}
	\item [1.] The first part is includes the $\times$2 scale methods (solid line) located in the 2 to 4 score range. BSRGAN and SwinIR of the $\times$2 scale showed better results compared to other methods, improving the SR image quality in terms of human perceived quality.
	\item [2.] The second part includes the $\times$2 scale methods located in the -1 to 2 score range. These methods recognized the low-quality image as the original and tried to enhance it as it was, resulting in amplified distortion and lower MOS scores
	\item [3.] The third part includes the $\times$4 scale methods (dotted line) located in the -1 to 2 score range.
	BSRGAN and SwinIR at the $\times$4 scale maintained the MOS score by removing the blur and noise factors, although the performance of restoring detailed information decreased as the scaling factors increased. However, BSRGAN's performance is partially poor in the screen content group (21-30), showing a low MOS score.
	\item [4.] The last part includes the $\times$4 scale methods located in the -3 to -1 score range.
	As with the previous $\times$2 scale methods, they not only amplify the distortion of the original but also show a greater decline in performance compared to BSRGAN and SwinIR due to the increased scaling factor.
\end{itemize}

\begin{figure}[!t]
	\centering
	\subfloat[]{\includegraphics[width=0.45\linewidth]{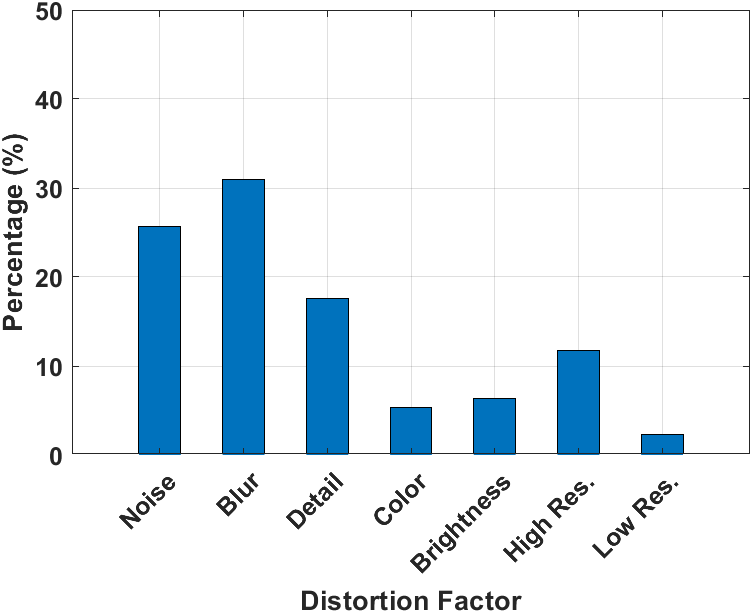}\label{fig:distfactor1}}\hspace{3mm}
	\subfloat[]{\includegraphics[width=0.45\linewidth]{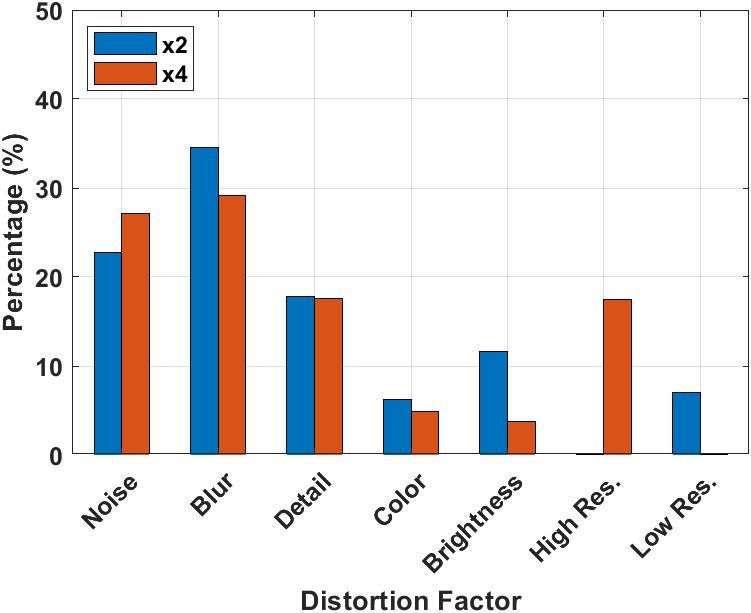}\label{fig:distfactor2}}
	\caption{Distributions of distortion for (a) Overall SR images and (b) based on scaling factor on SREB dataset.}
	\label{fig:distfactor}
	\vspace{-5mm}
\end{figure}

\subsubsection{Distortion Analysis}
\label{sec:DistAnalysis}
Based on the participants' questionnaire responses, we analyzed main distortion types, negatively affecting SR images quality. The distribution of distortion for all SR images is illustrated in Fig.~\ref{fig:distfactor}~\subref{fig:distfactor1}, with blur being the most dominant, followed by noise and loss of detail. Although SR typically induces blur rather than noise, noise from original low-quality broadcast content was found to be amplified in SR images, negatively affecting quality. Therefore, accurately evaluating blur and noise is critical for developing objective quality metrics for SR images from low-quality content. Fig.~\ref{fig:distfactor}~\subref{fig:distfactor2} represents the distribution of distortion based on scaling factors. Both scaling factors of $\times$2 and $\times$4 show that blur and noise are predominant, followed by detail.
The noise ratio increased in $\times$4 compared to $\times$2.
This is because the higher resolution allows for clearer identification of finer details in the image, thereby making noise easier to recognize.
Notably, in the $\times$4 scaling factor, high resolution was chosen as the fourth most frequent factor at 17.48\%.
This is due to existing distortions being amplified as the scale increases, resulting in an overall degradation in quality.

\begin{figure*}[!t]
	\centering
	\subfloat[]{\includegraphics[width=0.95\linewidth]{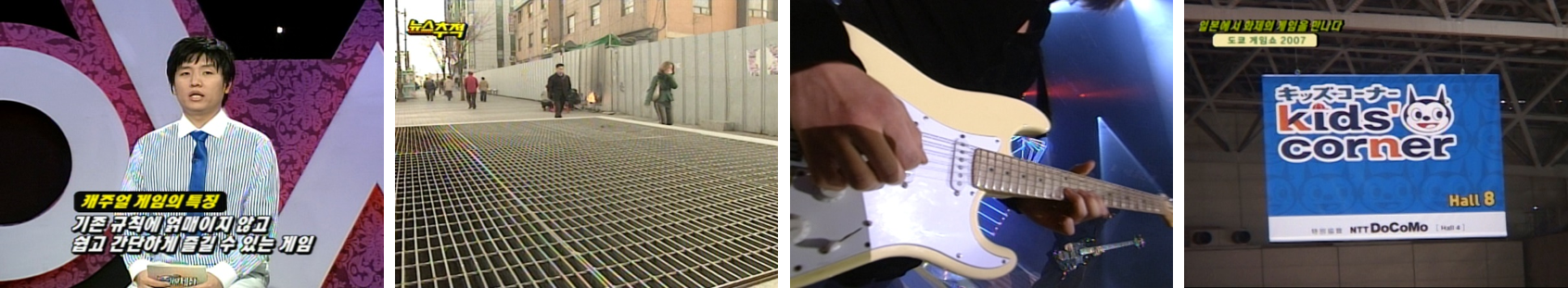}\label{fig:distimg1}} \\ \vspace{-3mm}
	\subfloat[]{\includegraphics[width=0.95\linewidth]{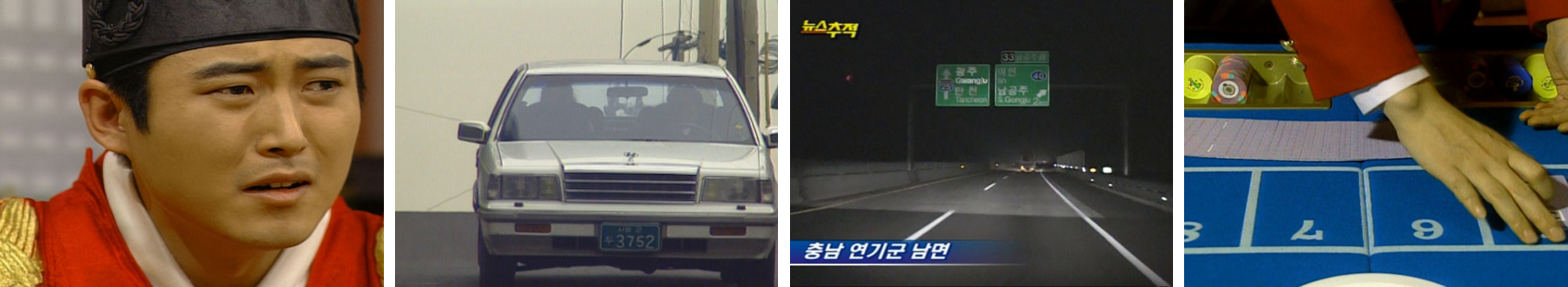}\label{fig:distimg2}}
	\caption{Contents dominated by (a) blur distortion and (b) noise distortion when SR is applied.}
	\label{fig:distimg}
	\vspace{-5mm}
\end{figure*}

Additionally, we conducted an analysis of distortions for each content. In Fig.~\ref{fig:distimg}~\subref{fig:distimg1}, it depicts instances where blur was predominantly selected as the primary distortion over noise. These images typically contain high levels of texture and edges, making noticeable areas of blur generated when SR is applied. The influence of blur on the sharpness of patterns and edges is evident, negatively affecting the overall image quality. Conversely, contents where noise was overwhelmingly chosen as the main distortion over blur are depicted in Fig.~\ref{fig:distimg}~\subref{fig:distimg2}. These images are often characterized by simplicity and a lack of texture, making noise more noticeable than blur within the images.

\section{Evaluation of Objective IQA Models}
\label{sec:evaluation}
In this section, we evaluate and compare the performance of existing IQA metrics on our SREB dataset.

\subsection{Test IQA Models \& Evaluation Metrics}
\label{sec:objectiveQA}
Since only low-resolution original images were available, we tested NR and RR IQA methods on the SREB dataset.
Specifically, nine NR IQA metrics were employed: BIQAA\cite{biqaa}, BRISQUE\cite{brisque}, BLIINDS II\cite{bliinds2}, SRmetric\cite{Ma}, HyperIQA\cite{hyperiqa}, TRes\cite{tres}, MANIQA\cite{maniqa}, KLTSRQA\cite{kltsrq}, and ARNIQA\cite{arniqa}. Additionally, two RR-IQA methods were employed: OSVP\cite{osvp}, REDLOG\cite{redlog}.

BIQAA \cite{biqaa} predicts the quality of images by measuring the variability in expected image entropy. BRISQUE \cite{brisque}, an NSS-based model, quantifies naturalness to predict quality scores using the Mean Subtracted Contrast Normalized (MSCN) coefficient along with Generalized Gaussian Distribution (GGD) and Asymmetric Generalized Gaussian Distribution (AGGD) calculations. BLIINDS II \cite{bliinds2} is a Bayesian inference model that predicts image quality using NSS models of Discrete Cosine Transform (DCT) coefficients, which enables simple learning and quality score prediction through feature extraction and probabilistic modeling. SRmetric \cite{Ma} combines local, global frequency, and spatial features using Principal Component Analysis (PCA) and steerable wavelet decomposition for quality prediction through regression modeling. HyperIQA \cite{hyperiqa} simulates how human evaluate image, enabling the network to recognize image content and adaptively adjust parameters for quality score prediction. TRes \cite{tres} integrates global and local features using CNNs and Transformers for feature extraction and score prediction. On the other hand, MANIQA \cite{maniqa} utilizes ViT for feature extraction and enhances the interaction between global and local image regions using modules that account for channel and spatial dimensions, predicting the final score with dual branches for each patch. KLTSRQA \cite{kltsrq}, a Karhunen-Loeve Transform (KLT)-based model, uses AGGD parameters and coefficient energy to extract image characteristics and predict quality. ARNIQA \cite{arniqa} introduces various distortions to pristine images to train the feature extractor on diverse distortion types, employing contrastive learning to enhance performance in image quality assessment by increasing correlation among identical distortions across different content.
One of the RR-IQA methods, OSVP\cite{osvp}, utilizes features that represent directional relationships with the surrounding regions of an image. It compares the features of a reduced reference image with those of a test image to calculate a quality score. REDLOG\cite{redlog} evaluates image quality by using the difference in entropy of Discrete Wavelet Transform (DWT) coefficients between a reduced reference image and a test image.

The objective of IQA metrics is to achieve high correlation with MOS. To evaluate the performance, we employed Pearson Linear Correlation Coefficient (PLCC), Spearman Rank Correlation Coefficient (SRCC), and Root Mean Square Error (RMSE), which represent prediction accuracy, monotonicity, and error, respectively. Higher correlation with MOS is indicated by PLCC and SRCC values closer to 1, and an RMSE closer to 0.

To conduct the evaluation, 5-fold cross-validation was utilized, with the SR images randomly split into 80\% for the training set and 20\% for the test set. Throughout 1000 iterations, we ensured random partitioning of the training and test sets in each iteration. For all experimental results obtained from these 1000 train-test repetitions, average values were used as final performance indices. Among the learning-based IQA models, those that extract features and utilize regressors adopted the Support Vector Regressor (SVR). ARNIQA utilized a pretrained encoder to extract features from the SREB dataset and trained a regressor using these features. To introduce nonlinearity to the predicted scores, we fitted them to MOS using the same non-linear logistic function when calculating PLCC and RMSE. The logistic function is defined as follows:
\begin{equation}
	f(x) = \frac{\lambda_1 - \lambda_2}{1 + \exp\left(\frac{-x + \lambda_3}{|\lambda_4|}\right)} + \lambda_2
	\label{eq:logistic}
\end{equation}
where $x$ represents the predicted score, $f(x)$ represents the fiited predicted score, and $\lambda_k~(k = 1, 2, 3, 4)$ are parameters aimed at minimizing the mean square error between $f(x)$ and MOS.
\begin{table*}[!t]
	\centering
	\caption{Test results for various IQA metrics on the SREB dataset}
	\label{tab:result}
	{\footnotesize
		\begin{tabular}{cc|ccc|ccc|ccc}
				\toprule
				\multicolumn{2}{c|}{\multirow{2}{*}{Method}}  & \multicolumn{3}{c|}{$\times$2}                                                                    & \multicolumn{3}{c|}{$\times$4}                                                                    & \multicolumn{3}{c}{$\times$2$\times$4}                                                                  \\ 
				& & {PLCC}         & {SRCC}         & RMSE         & {PLCC}         & {SRCC}         & RMSE             & {PLCC}           & {SRCC}         & RMSE         \\ 
				\midrule
				\multirow{9}{*}{FR} & BIQAA \cite{biqaa}                & {0.209}      & {0.241}        & 1.621          & {0.176}          & {0.174}          & 1.315          & {0.280}          & {0.358}        & 1.542          \\
				& BRISQUE \cite{brisque}            & {0.876}      & {0.805}        & 0.776          & {0.871}          & {0.714}          & \textbf{0.644} & {0.865}          & {0.820}        & 0.780          \\
				& BLIINDS II \cite{bliinds2}        & {0.820}      & {0.709}        & 0.921          & {0.848}          & {0.700}          & 0.686          & {0.779}          & {0.748}        & 0.974          \\
				& SRmetric \cite{Ma}                & {0.872}      & {0.793}        & 0.793          & {0.867}          & {0.725}          & 0.652          & {0.891}          & {0.828}        & 0.706          \\
				& HyperIQA \cite{hyperiqa}          & {0.738}      & {0.720}        & 1.574          & {0.720}          & {0.740}          & 1.288          & {0.802}          & {0.795}        & 1.094          \\
				& TRes \cite{tres}                  & {0.810}      & {0.781}        & 1.260          & {0.703}          & {0.653}          & 1.189          & {0.759}          & {0.748}        & 1.349          \\
				& MANIQA \cite{maniqa}              & {0.821}      & {0.664}        & 1.097          & {0.829}          & {0.710}          & 0.861          & {0.844}          & {0.808}        & 0.963          \\
				& KLTSRQA \cite{kltsrq}             & {0.910}      & {\textbf{0.861}} & \textbf{0.697} & {0.874}        & {0.724}          & 0.661          & {0.891}          & {0.858}        & 0.721          \\
				& ARNIQA \cite{arniqa}              & {\textbf{0.911}} & {0.860}      & 0.699        & {\textbf{0.878}} & {\textbf{0.775}} & 0.660          & {\textbf{0.932}} & {\textbf{0.899}} & \textbf{0.576} \\
				\midrule
				\multirow{2}{*}{RR} & OSVP \cite{osvp}                  & {0.669}      & {0.693}        & 1.639          & {0.673}          & {0.470}          & 1.321          & {0.178}          & {0.144}        & 1.565          \\
				& REDLOG \cite{redlog}              & {0.810}      & {0.734}        & 0.948          & {0.793}          & {0.616}          & 0.793          & {0.626}          & {0.576}        & 1.214          \\
				\bottomrule
			\end{tabular}
	}
	\vspace{-5mm}
\end{table*}

\subsection{Comparative Test Results}
\label{sec:comparativetest}
The evaluation results of the existing metrics on the SREB dataset are presented in Table \ref{tab:result}.
BIQAA and OSVP failed to accurately predict perceptual image quality.
These metrics consider the orientation of images, which can vary significantly depending on the content rather than the quality characteristics.
In addition, REDLOG failed to capture quality changes according to scaling factors, leading to significant performance degradation at x2x4 scaling factor.
Furthermore, BLIINDS II, HyperIQA, TRes, MANIQA did not exhibit robust performance in assessing SR images generated from low-quality broadcast content.
Even the recently proposed deep learning-based models such as HyperIQA, TRes, and MANIQA failed to achieve a PLCC of 0.85 or higher for all scaling factors. In contrast, the NSS-based model BRISQUE demonstrated PLCCs of at least 0.86, with the RMSE being the lowest at $\times$4 with a value of 0.644. SR-IQA models like SRmetric and KLTSRQA also exhibited good performance by achieving PLCCs of at least 0.86 across all scaling factors. 
Through consideration of SR, these two metrics demonstrated excellent performance.
Notably, KLTSRQA attained the best SRCC and RMSE of 0.861 and 0.697, respectively, at $\times$2. ARNIQA, utilizing deep learning-based degradation representation, demonstrated superior performance at $\times$4 and $\times$2$\times$4, achieving exceptional results with a PLCC of 0.932, a SRCC of 0.899, and a RMSE of 0.576 at $\times$2$\times$4. ARNIQA has been trained to effectively separate different types of distortion in manifold space, demonstrating the best performance on the SREB dataset.

\begin{figure}[!t]
	\centering
	\includegraphics[width=0.98\linewidth]{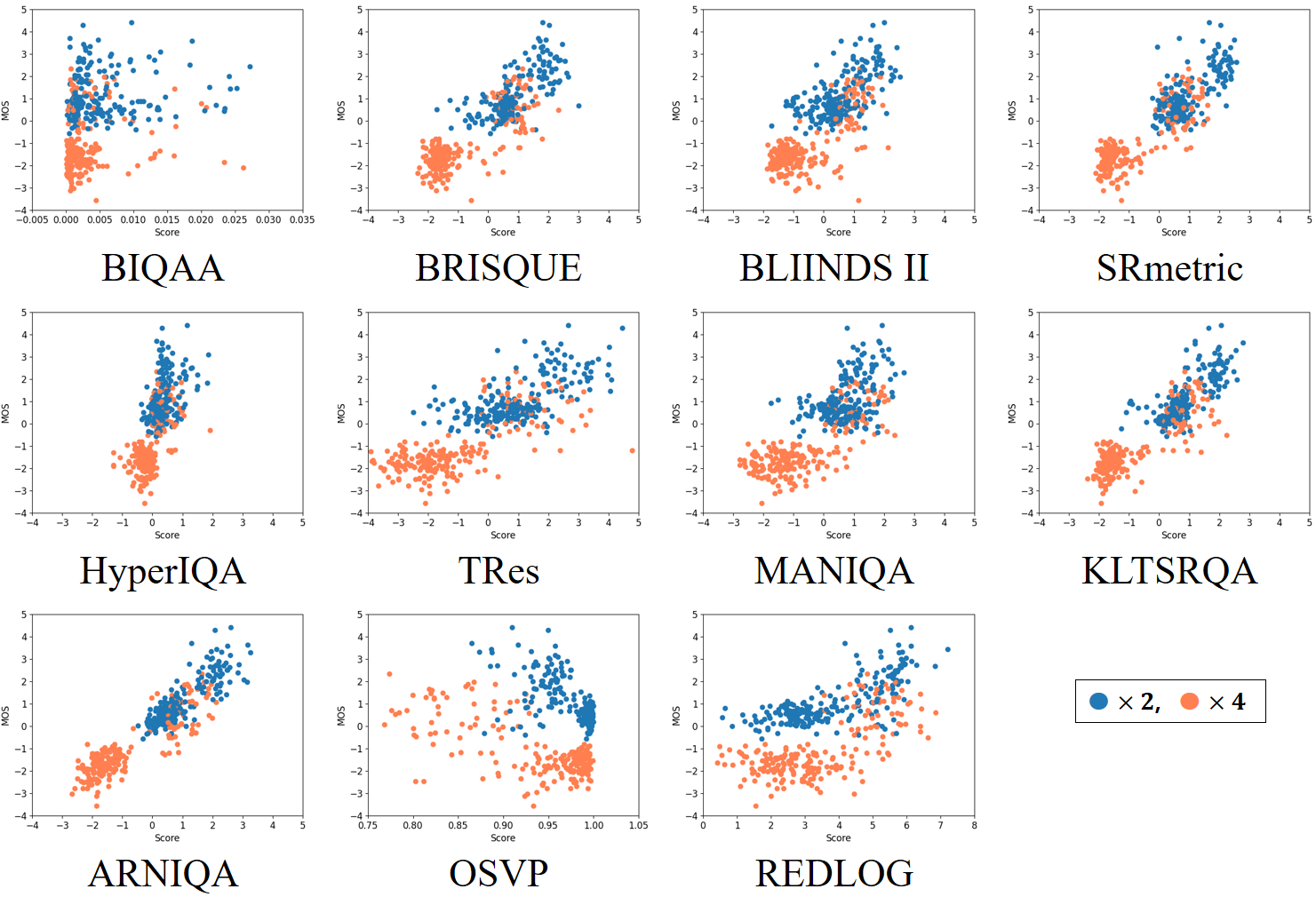}
	\caption{Scatter plots of predicted scores from various IQA metrics for $\times$2$\times$4 MOS on the SREB dataset.}
	\label{fig:2k4kscatter}
	\vspace{-5mm}
\end{figure}

\begin{figure}[!t]
	\centering
	\includegraphics[width=0.98\linewidth]{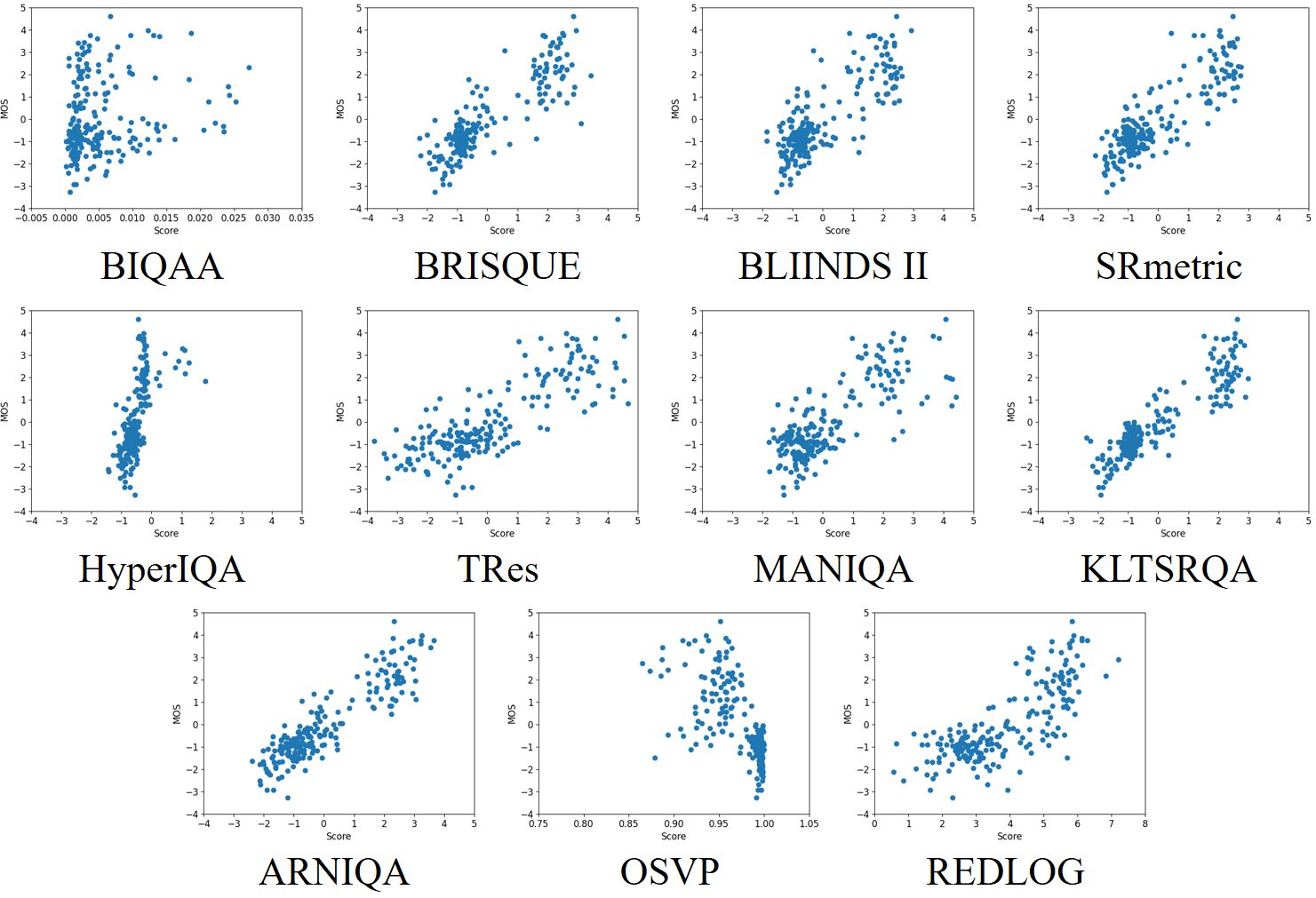}
	\caption{Scatter plots of predicted scores from various IQA metrics for $\times$2 MOS on the SREB dataset.}
	\label{fig:2kscatter}
	\vspace{-2mm}
\end{figure}

\begin{figure}[!t]
	\centering
	\includegraphics[width=0.98\linewidth]{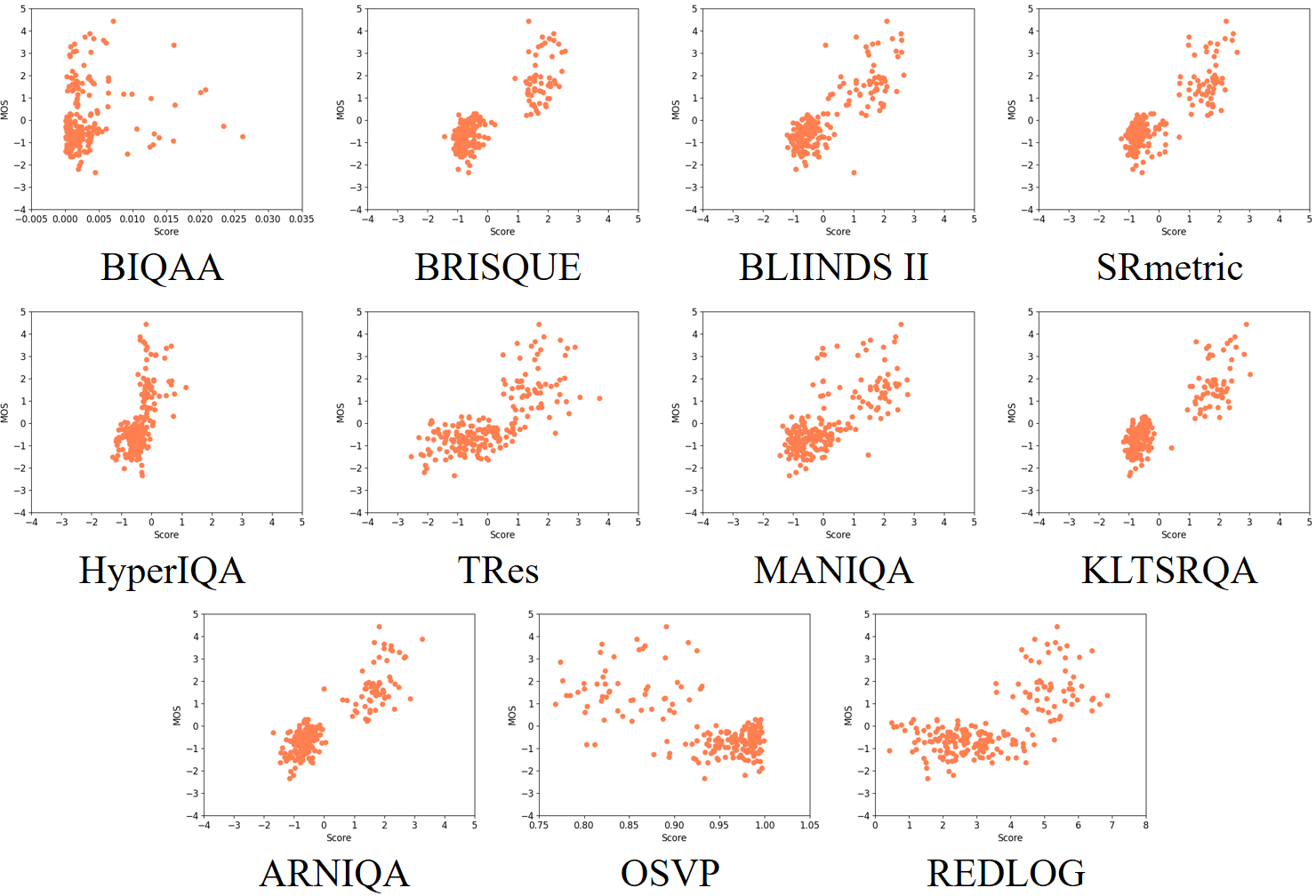}
	\caption{Scatter plots of predicted scores from various IQA metrics for $\times$4 MOS on the SREB dataset.}
	\label{fig:4kscatter}
	\vspace{-5mm}
\end{figure}

Fig.~\ref{fig:2k4kscatter} shows the scatter plots between $\times$2$\times$4 MOS of SREB and the predicted scores by each metric. 
BIQAA and OSVP show a widely scattered distribution due to its inconsistency with MOS.
In particular, OSVP shows a negative correlation between prediction scores and MOS.
This is because OSVP evaluates image quality based on the similarity between the SR image and the original image, and in the SREB dataset, high-quality SR images tend to have a greater difference from the low-quality original images.
REDLOG, along with BIQAA and OSVP, exhibits significantly degraded performance in the x2x4 scaling factor.
This is because these metrics failed to effectively capture scale changes, leading to nearly identical predictions for both x2 and x4 scaling factors.
The scatter plots of BLIINDS II, TRes, and MANIQA exhibit a positive correlation, but they are not tightly clustered, indicating a lower correlation. The HyperIQA exhibited a converged distribution, yet it showed a weak linear relationship standing vertically. Conversely, scatter plots of ARNIQA, KLTSRQA, and SRmetric depict tightly clustered distributions closely aligned along a line with a slope of 1, indicating a strong positive correlation between predicted scores and MOS.

Additionally, to investigate the performance differences of each metric based on scaling factors, we visualize scatter plots for $\times$2 and $\times$4 scaling factors and present them in Fig.~\ref{fig:2kscatter} and Fig.~\ref{fig:4kscatter}, respectively. They exhibited distributions similar to those of the $\times$2$\times$4 scatter plots, but discontinuities were observed. In particular, these discontinuities were more pronounced at $\times$4, resulting in a decrease in performance for most metrics at $\times$4 compared to $\times$2, as demonstrated in Table~\ref{tab:result}. To address this issue, future IQA metrics for SR applications must account for the increased distortion amplification resulting from resolution enhancement.

\section{Conclusion}
\label{sec:conclusion}
In this study, we conducted a human study to understand the perceptual quality of SR images generated from low-quality broadcast content. The proposed SR-IQA dataset, SREB, utilized original images at their native resolution without prior downsampling during the generation of SR images, considering real-world scenarios where SR is applied. 
SR methods that enhance or distort low-quality original images were used to generate SR images at 2K and 4K resolutions, applying two scaling factors ($\times$2, $\times$4) and seven different SR methods.

By conducting the subjective quality assessment test, we obtained human scores for SR images. Through MOS analysis, we confirmed that SwinIR produced the highest quality SR images, and perceptual quality decreased with increasing scaling factor across all SR methods. Furthermore, surveys revealed that primary distortions in SR images generated from low-quality broadcast content were blur,noise and loss of details.

We evaluated and compared existing IQA metrics on our dataset. In comparative experiments, ARNIQA demonstrated the highest correlation with MOS by considering various types and degrees of distortion using degradation representation. Additionally, we analyzed the limitations of existing models that performance degraded as the scaling factor increased.

The SREB dataset and insights from this study can be utilized to develop enhanced IQA metrics that account for scaling factor impacts as well as distortions and improvements commonly encountered in real-world SR applications.
Furthermore, these advanced IQA metrics can be applied in consumer technology, such as enabling smart photo editing apps to automatically improve image quality.

\bibliographystyle{IEEEtran}
\bibliography{Ref}

@inproceedings{SRCNN,
	  title={Learning a deep convolutional network for image super-resolution},
	  author={Dong, Chao and Loy, Chen Change and He, Kaiming and Tang, Xiaoou},
	  booktitle={Proc. Eur. Conf. Comput. Vis.},
	  year={2014},
	  pages={184--199},
}

@inproceedings{FSRCNN,
  	  title={Accelerating the super-resolution convolutional neural network},
  	  author={Dong, Chao and Loy, Chen Change and Tang, Xiaoou},
  	  booktitle={Proc. Eur. Conf. Comput. Vis.},
  	  pages={391--407},
  	  year={2016},
}

@inproceedings{VDSR,
  	  title={Accurate image super-resolution using very deep convolutional networks},
  	  author={Kim, Jiwon and Lee, Jung Kwon and Lee, Kyoung Mu},
  	  booktitle = {Proc. IEEE Conf. Comput. Vis. Pattern Recognit},
	  pages={1646--1654},
	  month = {June},
  	  year={2016}
}

@inproceedings{SRGAN,
  	  title={Photo-realistic single image super-resolution using a generative adversarial network},
  	  author={Ledig, Christian and others},
	  booktitle = {Proc. IEEE Conf. Comput. Vis. Pattern Recognit.},
  	  pages={4681--4690},
  	  year={2017}
}

@inproceedings{BSRGAN,
	  title={Designing a practical degradation model for deep blind image super-resolution},
	  author={Zhang, Kai and Liang, Jingyun and Van Gool, Luc and Timofte, Radu},
	  booktitle={Proc. IEEE Int. Conf. Comput. Vis.},
	  pages={4791--4800},
	  year={2021}
}

@inproceedings{SwinIR,
  	  title={Swinir: Image restoration using swin transformer},
  	  author={Liang, Jingyun and others},
  	  booktitle={Proc. IEEE Int. Conf. Comput. Vis.},
  	  pages={1833--1844},
  	  year={2021}
}

@ARTICLE{ssim,
	author={Zhou Wang and Bovik, A.C. and Sheikh, H.R. and Simoncelli, E.P.},
	journal={IEEE Transactions on Image Processing}, 
	title={Image quality assessment: from error visibility to structural similarity}, 
	year={2004},
	volume={13},
	number={4},
	pages={600-612},
}

@InProceedings{lplips,
	author = {Zhang, Richard and Isola, Phillip and Efros, Alexei A. and Shechtman, Eli and Wang, Oliver},
	title = {The Unreasonable Effectiveness of Deep Features as a Perceptual Metric},
	booktitle = {Proc. IEEE Conf. Comput. Vis. Pattern Recognit.},
	year = {2018}
}

@ARTICLE{redlog,
	author={Golestaneh, SeyedAlireza and Karam, Lina J.},
	journal={IEEE Transactions on Image Processing}, 
	title={Reduced-Reference Quality Assessment Based on the Entropy of DWT Coefficients of Locally Weighted Gradient Magnitudes}, 
	year={2016},
	volume={25},
	number={11},
	pages={5293-5303},
}

@article{osvp,
	author = {Jinjian Wu and Weisi Lin and Guangming Shi and Leida Li and Yuming Fang},
	journal = {Information Sciences},
	title = {Orientation selectivity based visual pattern for reduced-reference image quality assessment},
	year = {2016},
	volume = {351},
	pages = {18-29},
}

@article{igts,
	title = {A reduced-reference quality assessment metric for super-resolution reconstructed images with information gain and texture similarity},
	author = {Lijuan Tang and Kezheng Sun and Luping Liu and Guangcheng Wang and Yutao Liu},
	journal = {Signal Processing: Image Communication},
	volume = {79},
	pages = {32-39},
	year = {2019},
}

@ARTICLE{brisque,
	author={Mittal, Anish and Moorthy, Anush Krishna and Bovik, Alan Conrad},
	journal={IEEE Transactions on Image Processing}, 
	title={No-Reference Image Quality Assessment in the Spatial Domain}, 
	year={2012},
	volume={21},
	number={12},
	pages={4695-4708},
}

@ARTICLE{niqe,
	author={Mittal, Anish and Soundararajan, Rajiv and Bovik, Alan C.},
	journal={IEEE Signal Processing Letters}, 
	title={Making a “Completely Blind” Image Quality Analyzer}, 
	year={2013},
	volume={20},
	number={3},
	pages={209-212},
}

@ARTICLE{DIIVINE,
	author={Moorthy, Anush Krishna and Bovik, Alan Conrad},
	journal={IEEE Transactions on Image Processing}, 
	title={Blind Image Quality Assessment: From Natural Scene Statistics to Perceptual Quality}, 
	year={2011},
	volume={20},
	number={12},
	pages={3350-3364},
}

@article{ASDS,
  title={Image deblurring and super-resolution by adaptive sparse domain selection and adaptive regularization},
  author={Dong, Weisheng and Zhang, Lei and Shi, Guangming and Wu, Xiaolin},
  journal={IEEE Transactions on image processing},
  volume={20},
  number={7},
  pages={1838--1857},
  year={2011},
  publisher={IEEE}
}

@inproceedings{Set14,
  title={On single image scale-up using sparse-representations},
  author={Zeyde, Roman and Elad, Michael and Protter, Matan},
  booktitle={International Conference on Curves and Surfaces},
  pages={711--730},
  year={2010},
}

@article{LIVE,
  title={A statistical evaluation of recent full reference image quality assessment algorithms},
  author={Sheikh, Hamid R and Sabir, Muhammad F and Bovik, Alan C},
  journal={IEEE Transactions on image processing},
  volume={15},
  number={11},
  pages={3440--3451},
  year={2006},
  publisher={IEEE},
}

@ARTICLE{VDID2014,
	author={Gu, Ke and Liu, Min and Zhai, Guangtao and Yang, Xiaokang and Zhang, Wenjun},
	journal={IEEE Transactions on Broadcasting}, 
	title={Quality Assessment Considering Viewing Distance and Image Resolution}, 
	year={2015},
	volume={61},
	number={3},
	pages={520-531},
}

@ARTICLE{MARFD,
	author={Yue, Guanghui and Wu, Honglv and Yan, Weiqing and Zhou, Tianwei and Liu, Hantao and Zhou, Wei},
	journal={IEEE Transactions on Broadcasting}, 
	title={Subjective and Objective Quality Assessment of Multi-Attribute Retouched Face Images}, 
	year={2024},
	pages={1-15},
}

@inproceedings{Yang,
  title={Single-image super-resolution: A benchmark},
  author={Yang, Chih-Yuan and Ma, Chao and Yang, Ming-Hsuan},
  booktitle={European Conference on Computer Vision},
  pages={372--386},
  year={2014},
  organization={Springer}
}

@article{Waterloo,
  title={Objective quality assessment of interpolated natural images},
  author={Yeganeh, Hojatollah and Rostami, Mohammad and Wang, Zhou},
  journal={IEEE Transactions on Image Processing},
  volume={24},
  number={11},
  pages={4651--4663},
  year={2015},
  publisher={IEEE}
}

@article{Ma,
  title={Learning a no-reference quality metric for single-image super-resolution},
  author={Ma, Chao and Yang, Chih-Yuan and Yang, Xiaokang and Yang, Ming-Hsuan},
  journal={Computer Vision and Image Understanding},
  volume={158},
  pages={1-16},
  year={2017},
  publisher={Elsevier}
}

@article{sisrset,
	title = {SISRSet: Single image super-resolution subjective evaluation test and objective quality assessment},
	journal = {Neurocomputing},
	volume = {360},
	pages = {37-51},
	year = {2019},
	author = {Guangming Shi and Wenfei Wan and Jinjian Wu and Xuemei Xie and Weisheng Dong and Hong Ren Wu},
}

@ARTICLE{qads,
	author={Zhou, Fei and Yao, Rongguo and Liu, Bozhi and Qiu, Guoping},
	journal={IEEE Transactions on Image Processing}, 
	title={Visual Quality Assessment for Super-Resolved Images: Database and Method}, 
	year={2019},
	volume={28},
	number={7},
	pages={3528-3541},
}

@ARTICLE{kltsrq,
	author={Jiang, Qiuping and others},
	journal={IEEE Transactions on Image Processing}, 
	title={Single Image Super-Resolution Quality Assessment: A Real-World Dataset, Subjective Studies, and an Objective Metric}, 
	year={2022},
	volume={31},
	number={},
	pages={2279-2294},
}

@article{mdid,
	author = {Wen Sun and Fei Zhou and Qingmin Liao},
	journal = {Pattern Recognition},
	title = {{MDID}: A multiply distorted image database for image quality assessment},
	year = {2017},
	volume = {61},
	pages = {153-168},
}

@book{p910,
	title        	= {Subjective Video Quality Assessment Methods for Multimedia Applications},
	publisher		= {document ITU-T Rec. P.910},
	year         	= {2008},
	url		   	= {https://www.itu.int/rec/T-REC-P.910},
}

@inproceedings{cf,
	title={Measuring colorfulness in natural images},
	author={Hasler, David and Suesstrunk, Sabine E},
	booktitle={Human vision and electronic imaging VIII},
	volume={5007},
	pages={87--95},
	year={2003},
	organization={SPIE}
}

@book{bt500,
    title 		= {Methodologies for the Subjective Assessment of the Quality of Television Images},
    publisher 	= {document ITU-R Rec. BT.500},
    year 		= {2019},
    url 		= {https://www.itu.int/rec/R-REC-BT.500},
}

@article{btmodel,
	title={Rank analysis of incomplete block designs: I. The method of paired comparisons},
	author={Bradley, Ralph Allan and Terry, Milton E},
	journal={Biometrika},
	volume={39},
	number={3},
	pages={324-345},
	year={1952},
}

@article{biqaa,
	author = {Salvador Gabarda and Gabriel Crist\'{o}bal},
	journal = {J. Opt. Soc. Am. A},
	number = {12},
	pages = {B42--B51},
	publisher = {Optica Publishing Group},
	title = {Blind image quality assessment through anisotropy},
	volume = {24},
	year = {2007},
}

@ARTICLE{bliinds2,
	author={Saad, Michele A. and Bovik, Alan C. and Charrier, Christophe},
	journal={IEEE Transactions on Image Processing}, 
	title={Blind Image Quality Assessment: A Natural Scene Statistics Approach in the DCT Domain}, 
	year={2012},
	volume={21},
	number={8},
	pages={3339-3352},
}

@InProceedings{hyperiqa,
	author = {Su, Shaolin and others},
	title = {Blindly Assess Image Quality in the Wild Guided by a Self-Adaptive Hyper Network},
	booktitle = {Proceedings of the IEEE/CVF Conference on Computer Vision and Pattern Recognition (CVPR)},
	year = {2020},
}

@InProceedings{tres,
    author    = {Golestaneh, S. Alireza and Dadsetan, Saba and Kitani, Kris M.},
    title     = {No-Reference Image Quality Assessment via Transformers, Relative Ranking, and Self-Consistency},
    booktitle = {Proceedings of the IEEE/CVF Winter Conference on Applications of Computer Vision (WACV)},
    month     = {January},
    year      = {2022},
    pages     = {1220-1230},
}

@InProceedings{maniqa,
    author    = {Yang, Sidi and others},
    title     = {{MANIQA}: Multi-Dimension Attention Network for No-Reference Image Quality Assessment},
    booktitle = {Proceedings of the IEEE/CVF Conference on Computer Vision and Pattern Recognition (CVPR) Workshops},
    year      = {2022},
    pages     = {1191-1200},
}

@InProceedings{arniqa,
    author    = {Agnolucci, Lorenzo and Galteri, Leonardo and Bertini, Marco and Del Bimbo, Alberto},
    title     = {{ARNIQA}: Learning Distortion Manifold for Image Quality Assessment},
    booktitle = {Proceedings of the IEEE/CVF Winter Conference on Applications of Computer Vision (WACV)},
    year      = {2024},
    pages     = {189-198}
}

@article{camera,
  title={Subjective image quality assessment based on objective image quality measurement factors},
  author={Park, Hyung-Ju and Har, Dong-Hwan},
  journal={IEEE Transactions on Consumer Electronics},
  volume={57},
  number={3},
  pages={1176--1184},
  year={2011},
  publisher={IEEE}
}

@article{HEVC,
  title={Subjective evaluation of HEVC and AVC/H. 264 in mobile environments},
  author={Garcia, Ray and Kalva, Hari},
  journal={IEEE Transactions on Consumer Electronics},
  volume={60},
  number={1},
  pages={116--123},
  year={2014},
  publisher={IEEE}
}

\vspace{11pt}

\begin{IEEEbiography}[{\includegraphics[width=1in,height=1.25in,clip,keepaspectratio]{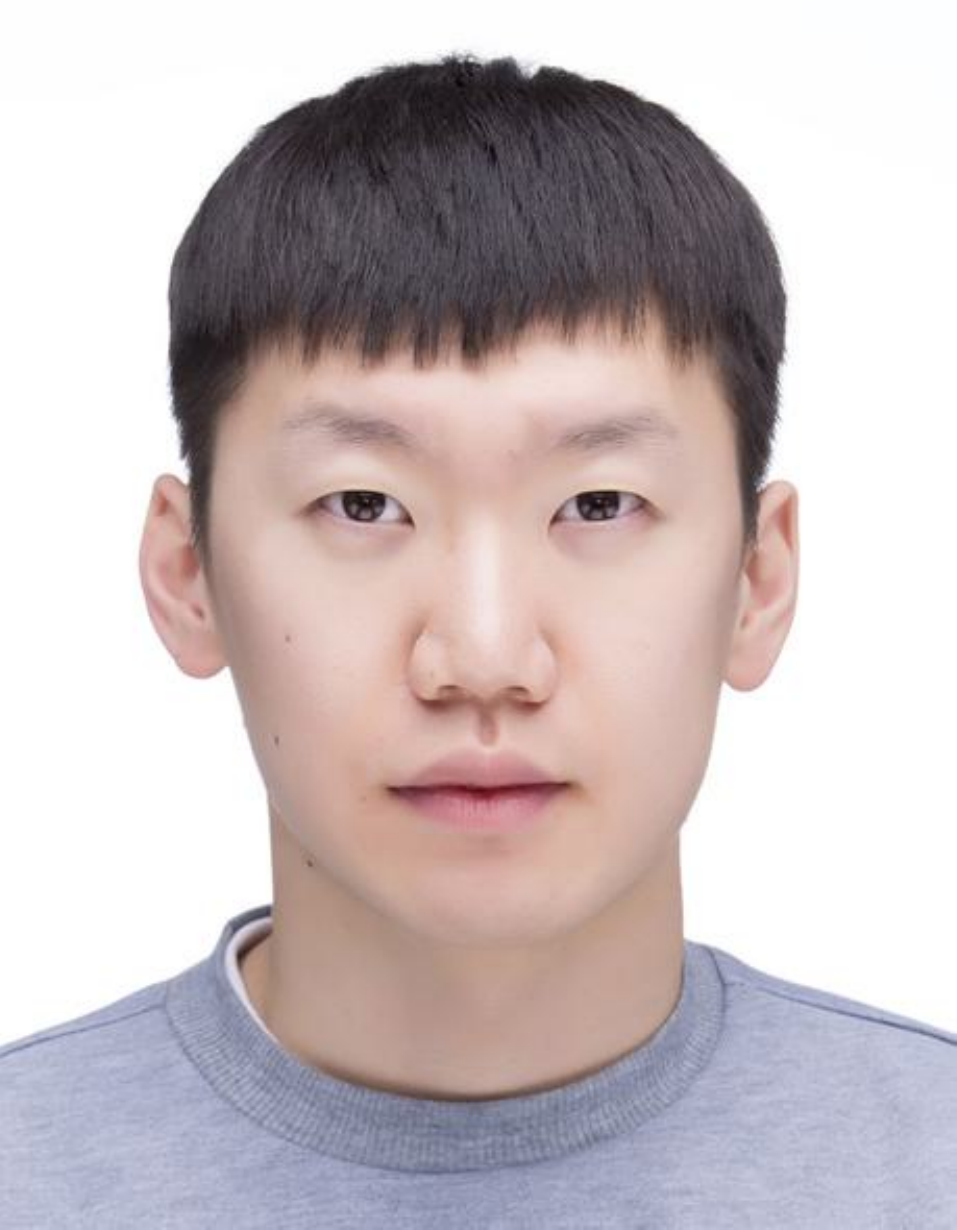}}]{Yongrok Kim}
	received the B.S. degree from the School of Electrical Engineering, Hanyang University ERICA, Ansan, South Korea, in 2020, where he is currently pursuing the integrated M.S. and Ph.D. degrees from the Department of Electrical and Electronic Engineering. His research interests are in the areas of visual quality assessment and deep learning-based image restoration.
\end{IEEEbiography}

\begin{IEEEbiography}[{\includegraphics[width=1in,height=1.25in,clip,keepaspectratio]{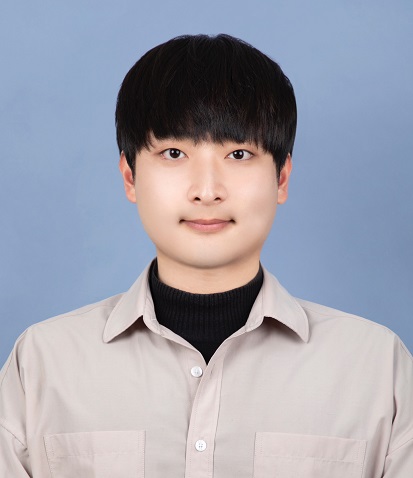}}]{Junha Shin}
	received the B.S. degree from the School of Electrical Engineering, Hanyang University ERICA, Ansan, South Korea, in 2023, where he is currently pursuing the integrated M.S. and Ph.D. degrees from the Department of Electrical and Electronic Engineering. His research interests are in the areas of perceptual visual quality assessment, deep learning based image quality assessment and generative model.
\end{IEEEbiography}

\begin{IEEEbiography}[{\includegraphics[width=1in,height=1.25in,clip,keepaspectratio]{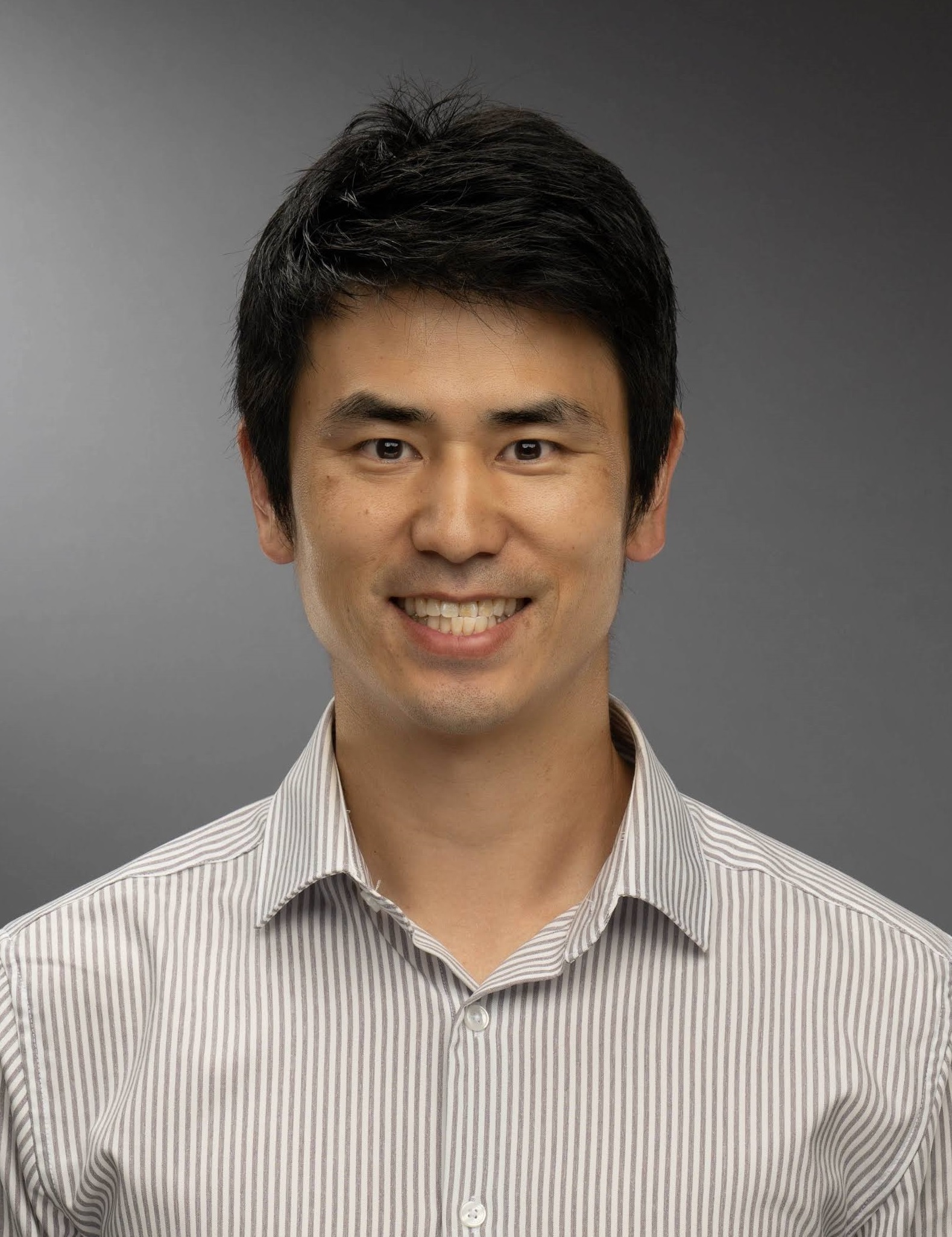}}]{Juhyun Lee}
	currently holds the position of Associate Professor in the Department of Bioengineering at the University of Texas at Arlington (UTA). He received a Bachelor of Science (B.S.) in Biomedical Engineering from the University of Utah in 2010. Subsequently, Dr. Lee pursued his graduate studies, earning a Master's degree from the University of Southern California (USC) and a Ph.D. in Bioengineering from the University of California, Los Angeles (UCLA). Following his academic training, he briefly transitioned to the industry, working as a biomedical engineer at Edward Lifesciences. After coming back to academia, Dr. Lee has been dedicated to advancing the field of high-resolution microscopy for in vivo dynamic imaging techniques. 
\end{IEEEbiography}

\begin{IEEEbiography}[{\includegraphics[width=1in,height=1.25in,clip,keepaspectratio]{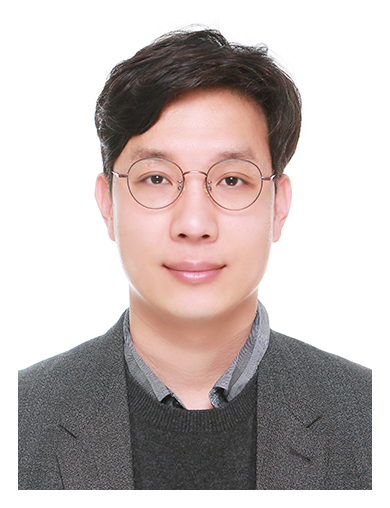}}]{Hyunsuk Ko}
	(Member, IEEE) received the B.S. and M.S. degrees from the Department of Electrical Engineering, Yonsei University, Seoul, South Korea, in 2006 and 2009, respectively, and the Ph.D. degree from the Department of Electrical Engineering, University of Southern California, Los Angeles, CA, USA, in 2015. From 2015 to 2020, he had been a Senior Researcher in the Electronics and Telecommunications Research Institute. He is currently an Associate Professor with the School of Electrical Engineering, Hanyang University ERICA, Ansan, South Korea. His research interests are in the areas of video coding, perceptual visual quality assessment, computer vision and deep learning-based multimedia signal processing.
\end{IEEEbiography}

\vfill

\end{document}